\pdfoutput=1
\documentclass[preprint,12pt]{elsarticle}

\usepackage{lmodern}
\usepackage{amsmath}
\usepackage{graphicx}
\usepackage{booktabs}
\usepackage{float}
\usepackage{subcaption}
\usepackage{hyperref}

\journal{Online Social Networks and Media}

\begin{document}

\begin{frontmatter}

\title{Predicting, Evaluating, and Explaining Top Misinformation Spreaders
via Archetypal User Behavior}

\tnotetext[vor]{This is the authors' version of a work published in
\emph{Online Social Networks and Media} \textbf{50} (2025) 100336,
\url{https://doi.org/10.1016/j.osnem.2025.100336}.
\textcopyright{} 2025 The Authors. Published by Elsevier B.V. This is an open access
article distributed under the terms of the Creative Commons Attribution 4.0
International (CC BY 4.0) licence, \url{https://creativecommons.org/licenses/by/4.0/}.}

\author[label1]{Enrico Verdolotti}

\author[label2]{Luca Luceri}

\author[label1]{Silvia Giordano}

\affiliation[label1]{
organization={ISIN - DTI, SUPSI},
city={Lugano},
country={Switzerland}
}

\affiliation[label2]{
organization={USC Information Sciences Institute},
city={Marina del Rey, California},
country={USA}
}

\begin{abstract}
The spread of misinformation on social networks poses a significant challenge to online communities and society at large. Not all users contribute equally to this phenomenon: a small number of highly effective individuals can exert outsized influence, amplifying false narratives and contributing to significant societal harm. This paper seeks to mitigate the spread of misinformation by enabling proactive interventions, identifying and ranking users according to key behavioral indicators associated with harmful content dissemination. We examine three user archetypes---\emph{amplifiers}, \emph{super-spreaders}, and \emph{coordinated accounts}---each characterized by distinct behavioral patterns in the dissemination of misinformation. These are not mutually exclusive, and individual users may exhibit characteristics of multiple archetypes. We develop and evaluate several user ranking models, each aligned with a specific archetype, and find that \emph{super-spreader} traits consistently dominate the top ranks among the most influential misinformation spreaders. As we move down the ranking, however, the interplay of multiple archetypes becomes more prominent. Additionally, we demonstrate the critical role of temporal dynamics in predictive performance, and introduce methods that reduce data requirements by minimizing the observation window needed for accurate forecasting. Finally, we demonstrate the utility and benefits of explainable AI (XAI) techniques, integrating multiple archetypal traits into a unified model to enhance interpretability and offer deeper insight into the key factors driving misinformation propagation. Our findings provide actionable tools for identifying potentially harmful users and guiding content moderation strategies, enabling platforms to monitor accounts of concern more effectively. 
\end{abstract}

\begin{graphicalabstract}
\includegraphics{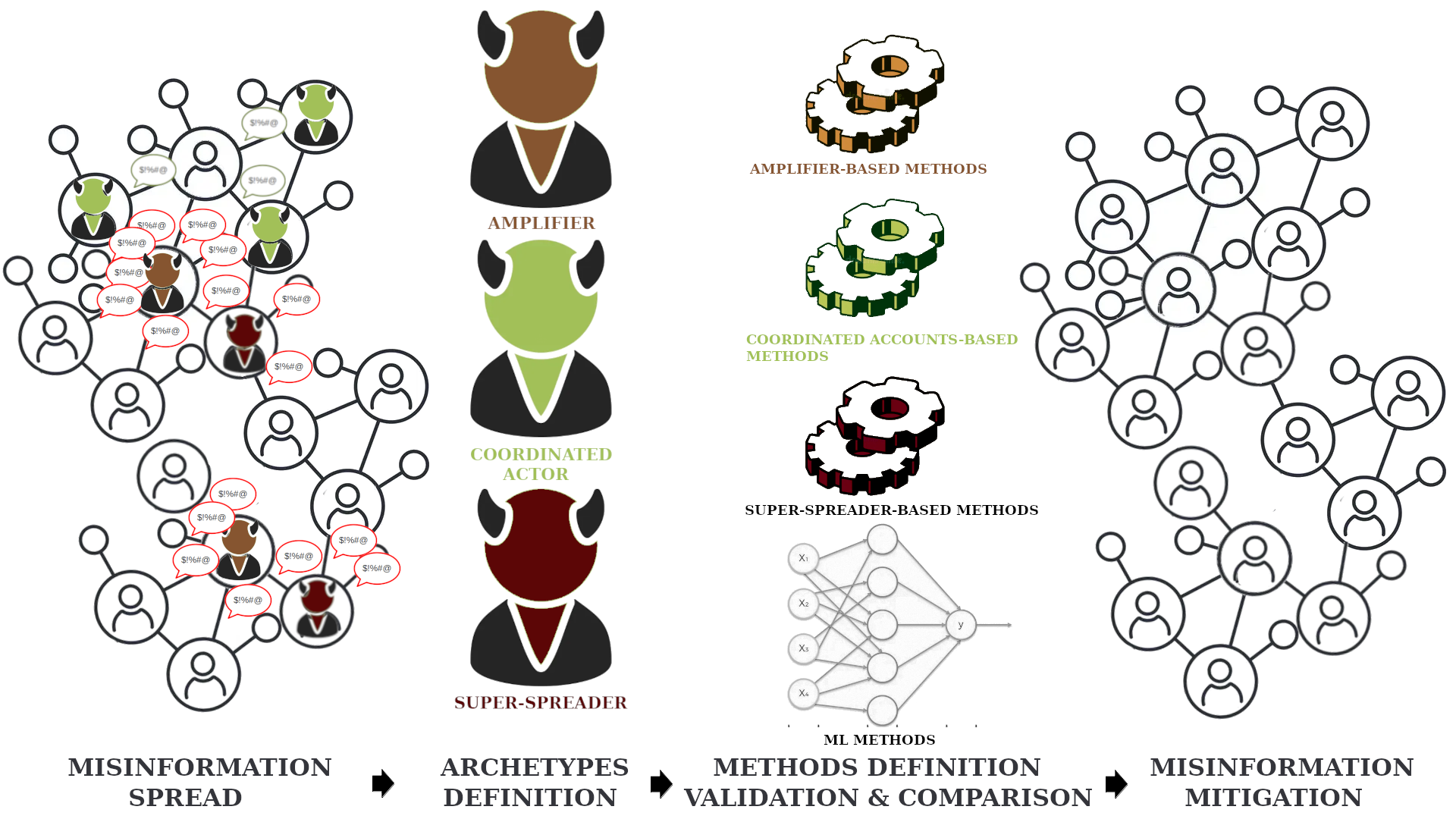}
\end{graphicalabstract}

\begin{highlights}
\item Formalization and refinement of user behavioral archetypes: \emph{amplifiers}, \emph{super-spreaders}, and \emph{coordinated accounts}.
\item Development, evaluation, and comparison of several methods based on distinct archetypes, and their combination, for predicting top misinformation spreaders.
\item Development of a methodology based on explainable AI to provide insights on the diverse behavioral traits of the actors mainly responsible for misinformation propagation.
\end{highlights}

\begin{keyword}
social media \sep misinformation \sep prediction \sep prevention \sep user behavior \sep content moderation \sep network analysis \sep early warning

\end{keyword}

\end{frontmatter}

\section{Introduction}\label{section:introduction}

Social media platforms have transformed the way information circulates, enabling billions of users to create, share, and consume content on an unprecedented scale. This interconnected digital environment encourages open communication and civic engagement but also facilitates the rapid spread of misleading claims and misinformation~\cite{bhattacharya2025unravelling}. During the COVID-19 pandemic, for example, false and low-credibility content proliferated across platforms~\cite{Pierri_2023,verma2022examining,Nogara2022TheDD}, shaping public perception and leading to harmful behaviors~\cite{Cinelli_2020,cheng2021twittervsfacebook,gatta2023interconnected,gisondi2022deadly}. The consequences of such misinformation are far-reaching: it can distort individual and collective decision-making, polarize public discourse, and erode trust in science, media, and democratic institutions~\cite{Erisen2025Trust,luceri2021social}.

Despite growing awareness of the dangers posed by misinformation, most moderation strategies employed by social media platforms remain fundamentally reactive. Interventions are often triggered only after a post has been flagged by users, detected by automated filters, or reported by fact-checkers, typically well after the content has begun to circulate widely~\cite{edelson2025measurement}. By the time action is taken, such as labeling, demotion, or removal, the harmful information may have already reached large audiences or shaped public opinion~\cite{pennycook2021shifting,truong_delayed_2025}. This lag in response significantly constrains the effectiveness of moderation efforts, as it allows false or misleading narratives to gain legitimacy, exploit algorithmic amplification~\cite{meerson2025platform}, and embed themselves in the public consciousness. Moreover, the viral nature of misinformation means that even short delays can have outsized consequences. Studies have shown that false content often spreads faster and more broadly than factual corrections, which struggle to catch up once a narrative has taken root~\cite{truong_delayed_2025,vosoughi2018spread, solovev2022moralemotionsshapevirality}.

Bridging this gap calls for proactive, scalable solutions that can detect and counteract misinformation before it gains widespread traction. In this context, a pressing research question arises: \textit{Can we estimate the potential of an account to disseminate misinformation based on specific behavioral traits—or, put differently, does an account’s behavioral footprint reveal latent signals of misinformation risk?} Exploring this question could pave the way for early-warning systems that identify high-risk users before false content circulates broadly. By enabling timely, targeted interventions, such approaches have the potential to significantly enhance moderation efforts and reduce the overall harm caused by misinformation proliferation. To explore this research question, we examine three key behavioral archetypes that may characterize accounts responsible for the dissemination of misinformation:
\textit{(i) Amplifiers}: These accounts generate little to no original content but significantly boost the reach of information by re-sharing posts from others~\cite{Wang2022IdentifyingAC}. Their defining trait is a high frequency of re-sharing activity.
\textit{(ii) Super-spreaders}: Highly influential users whose original content consistently goes viral~\cite{deverna_identifying_2024,baribi2024supersharers}. They are characterized by the volume (and virality) of the content they create and the high levels of engagement they receive from other users.
\textit{(iii) Coordinated accounts}: Groups of users who collaborate to promote specific narratives by artificially amplifying targeted content~\cite{pacheco_uncovering_2021,luceri2024unmasking}. These accounts typically exhibit synchronized posting behavior and strong behavioral similarity through their orchestrated actions~\cite{tardelli2024temporal,cinus2025exposing}.

To support more effective mitigation of misinformation, we build on these archetypes by developing and evaluating several user ranking systems that assess users based on their potential to spread misinformation. Each system is centered on a specific behavioral trait corresponding to one of the archetypes. In addition, we introduce a machine learning–based ranking framework that integrates all three archetypes by using their associated traits as input features. This approach serves two primary objectives: first, to evaluate whether integrating multiple behavioral signals improves predictive performance; and second, to apply explainable AI techniques to determine which traits most significantly influence a user’s ranking.

Our results reveal that the \emph{super-spreader} archetype dominates the highest-ranked positions---those most responsible for misinformation dissemination---showing that a small number of content creators play a disproportionate role in the spread of false information, in line with previous findings~\cite{Nogara2022TheDD,deverna_identifying_2024,baribi2024supersharers}.

In contrast, at middle and lower ranks, multiple archetypal traits frequently co-occur within individual accounts involved in the spread of misinformation. For example, resharing behavior—a defining characteristic of \textit{amplifier} accounts—becomes increasingly prevalent further down the ranking. This trend suggests that while super-spreaders are instrumental in initiating virality, \textit{amplifiers} and \textit{coordinated accounts} play a substantial role in sustaining and extending the reach of misinformation across the network.

This paper offers three key contributions.
First, it formalizes and refines the behavioral archetypes of misinformation actors, addressing a critical gap in the literature where terms like \textit{amplifiers} remain loosely defined, and research has predominantly focused on \textit{super-spreaders }and \textit{coordinated accounts}~\cite{pacheco_uncovering_2021, sharma2021identifyingcoordinatedaccountssocial, ng2023coordinatedinformationcampaignssocial, chen2025synthetic, pantè2025interactionpatternsassessingclaims}.
Second, it underscores the importance of temporal dynamics in misinformation diffusion and introduces the Time-Aware Social H-index (TASH-Index)—a novel, computationally efficient ranking method that performs comparably to more sophisticated approaches while requiring fewer resources.
Third, it presents a machine learning-based methodology that integrates features from multiple archetypes and leverages explainable AI techniques to illuminate the behavioral signals driving misinformation propagation. This approach provides a more nuanced understanding of how different user behaviors—beyond content generation—contribute to the spread of low-credibility content in networked environments.

Together, these contributions provide actionable tools for identifying potentially harmful users and informing content moderation practices. By surfacing influential spreaders early and shedding light on their behavioral profiles, our approach enables more efficient monitoring and prioritization of accounts of concern. The integration of explainable AI further enhances transparency and interpretability, offering a foundation for future interdisciplinary research that bridges computational methods and social science to better understand and mitigate the spread of misinformation at scale.

\section{Related Work}\label{section:related_work}

\subsection{Misinformation Detection: Content vs. User-centered Approaches}

The study of misinformation detection has evolved along two main perspectives: content-centered and user-centered. These approaches differ in their conceptual framing of the problem and in the methodologies they employ. This section reviews both perspectives, highlighting their core contributions and limitations, and positioning our work within this landscape.

Content-centered research focuses on analyzing the characteristics of the information being shared. It seeks to answer questions such as: \emph{Is this content misinformation?} \emph{What features make it likely to go viral?} and \emph{Which modalities contribute to its diffusion?} Early methods relied on lexical analysis, sentiment detection, and simple keyword-based heuristics. More recent advances use deep learning models, including LSTMs, attention mechanisms, and transformers, to detect linguistic or semantic patterns indicative of misinformation~\cite{shaeri_semi-supervised_2024, guo_tiefake_2023, karim_strengthening_2024}. In multimodal settings, researchers have combined textual and visual signals to capture richer representations of content virality~\cite{abdali_multi-modal_2024}. Recently, the diffusion of AI-generated content has drawn comparisons to the spread of misinformation, revealing notable similarities that warrant attention \cite{chen2025synthetic}. In both cases, a small number of highly visible actors, such as verified accounts and automated bots, are responsible for a disproportionately large share of the content’s propagation.

While content-centered approaches sometimes incorporate user-related features, such as the identity of the author or engagement metrics, the main analytical focus remains on the content itself. Consequently, they often overlook the role of user behavior and network dynamics in shaping diffusion patterns. In contrast, user-centered approaches emphasize the behaviors, attributes, and network positions of individuals participating in the spread of misinformation. These studies focus on identifying key actors, analyzing activity patterns, and modeling interactions within social networks. Rather than treating users as auxiliary features, user-centered models seek to explain diffusion phenomena through user behavior itself.

Our work builds on this user-centered perspective, with the objective of modeling misinformation diffusion as the emergent outcome of user activity patterns. We propose a framework that focuses on \emph{behavioral archetypes} rather than content features. This represents a key contribution of our work and will be explored in greater depth in the following sections.

\subsection{User-Centered Approaches for Misinformation Detection}

Research grounded in the user-centered paradigm has examined a wide range of behavioral dimensions to better understand misinformation dynamics. Prior studies have analyzed temporal activity patterns~\cite{stockinger_connection_2023}, network centrality and influence metrics~\cite{zhou_beyond_2024}, and exposure patterns linked to misinformation risk~\cite{avram_exposure_2020}. Other studies have investigated ideological alignment, hate speech, network exposure, and psychological traits, such as cognitive reflection and susceptibility to influence, as key factors underlying the adoption of misinformation~\cite{karami_profiling_2021, ye2024susceptibility, luceri2025susceptibility}. For example, Ye et al.~\cite{ye2024susceptibility} found that social media users require fewer exposures to adopt low-credibility content compared to high-credibility content. 

From a methodological standpoint, several efforts have leveraged graph neural networks (GNNs) to capture relational structures in social media data. Rath et al.~\cite{rath_detecting_2020} introduced an inductive GNN-based approach for detecting fake news spreaders, which was later extended with attention mechanisms and explanation modules in SCARLET~\cite{rath_scarlet_2021}. Sakketou et al.~\cite{sakketou_temporal_2022} incorporated temporal dynamics into GNNs, enabling forecasts based on evolving behavior. Ullah et al.~\cite{ullah_identifying_2023} benchmarked different GNN architectures, discussing their trade-offs between expressiveness and scalability. Minici et al.~\cite{minici2025iohunter} leveraged a GNN to model similarity networks based on users’ sharing activity and detect coordinated accounts driving information operations.

Beyond GNN-based approaches, research in the misinformation domain has advanced along three main directions, each targeting a distinct objective to deepen our understanding of misinformation dynamics: 
(i) identifying active spreaders through heuristics or supervised models; 
(ii) predicting future spreaders using temporal and behavioral predictors; and 
(iii) characterizing user roles and strategies based on ideological, linguistic, or network features. 

Despite these advances, several critical challenges remain. In particular, the distinct roles of user archetypes in misinformation diffusion—especially the emergence and impact of amplifiers, who propagate but do not originate false content—remains underexplored. Additionally, user roles tend to be treated as static categories, despite evidence of dynamic shifts over time. Finally, as models grow more complex, issues of transparency and explainability become critical, especially when findings are intended to inform policy decisions. While explainable AI (XAI) techniques are emerging in this domain, their adoption in misinformation research is still limited.

Our approach contributes to this area of research in two ways. First, we introduce a formalization of three distinct behavioral archetypes that characterize users involved in the spread of misinformation. Second, we enhance a previously proposed influence metric—an H-index-based score introduced by \cite{deverna_identifying_2024}—by making it time-aware, thus capturing temporal patterns in user activity. This refined metric, along with other archetype-specific scores, is integrated into a machine learning framework designed to estimate a user’s potential to disseminate misinformation. By combining features derived from multiple archetypes, the model captures interdependencies between behavioral traits while maintaining interpretability. Leveraging SHAP-based explainability methods, our approach enables the characterization of individual users in terms of their dominant archetypal behaviors. This not only provides deeper insight into the dynamics of misinformation diffusion but also offers actionable evidence that may inform the development of mitigation strategies and platform governance policies.

\section{Behavioral Archetypes: Definition and Characterization}\label{sec:behavioral_archetypes}  

In this study, we delineate three key archetypes of user behavior on social media: \emph{amplifiers}, \emph{super-spreaders}, and \emph{coordinated accounts}. These archetypes are not mutually exclusive, as individual users may exhibit characteristics of multiple types. They encompass both human-operated and automated accounts, such as social bots \cite{ferrara2016rise,luceri2019red}, and constitute the foundation for the ranking models introduced in the subsequent sections. We define each archetype below, based exclusively on its content diffusion behavior:

\begin{itemize}

    \item \emph{Amplifiers} are users whose primary strategy for spreading information consists of extensive resharing. While the term ``amplifier'' is often used in the literature to describe broader platform-level phenomena~\cite{Lim2022OpinionAC, Nowak-Teter2024, Peck2020APO}, similarly to~\cite{Wang2022IdentifyingAC}, we employ it here to characterize accounts that systematically amplify content by re-sharing posts, rather than by creating original information.
    
    \item \emph{Coordinated accounts} represent groups of users that operate in collaborative fashion, often driving orchestrated campaigns. These accounts exhibit behaviors reminiscent of swarm-like coordination, frequently amplifying the same messages to elicit the illusion of public consensus. Coordination on social media has been widely documented~\cite{Hristakieva2021TheSO, Sharma2020IdentifyingCA,vishnuprasad2024tracking}, with evidence suggesting its employment in multiple influence campaigns aimed at manipulating public opinion and recommendation algorithms~\cite{pacheco_uncovering_2021,luceri2024unmasking,luceri2025coordinated}, even across platforms \cite{minici2024uncovering,cinus2025exposing}.
    
    \item \emph{Super-spreaders} are users who consistently generate and share original content that achieves viral reach through extensive re-sharing by others on social media platforms. 
    These users play a critical role in initiating and fueling the diffusion of narratives. For this reason, they are particularly influential in misinformation dynamics~\cite{deverna_identifying_2024}.
    
\end{itemize}

These three behavioral archetypes encapsulate distinct yet interrelated aspects of user activity in information spread. In the following sections, we formalize methodologies for identifying and ranking users who exhibit these behaviors in the diffusion of misinformation. Our models assign each account a score reflecting its potential to propagate low-credibility content, enabling a structured approach to assessing online risks and uncovering its underlying drivers.

\section{Methodology}\label{section:methodology}

In this section, we present the approach to establishing content credibility, and we introduce the ranking methodologies we adopt for analysis, including both archetype-based and machine learning models.

\subsection{Credibility Assessment}\label{subsec:credibility_assessment}

To assess the credibility of social media content, we adopt a domain-based approach that infers credibility from the trustworthiness of URLs embedded in user posts. Following established practice~\cite{deverna_identifying_2024}, we consider only original posts and reshares, excluding self-reshares, replies, and quotes, as these do not reflect content endorsement or diffusion. Posts are labeled using NewsGuard’s\footnote{\href{www.newsguardtech.com}{www.newsguardtech.com}} credibility scores, which assign trust ratings in the range [0--100] to news domains. This platform-agnostic method ensures consistency with prior work and supports scalable, interpretable credibility assessments. Following NewsGuard guidelines, we categorize posts linking to domains with a score of 39 or lower as \textit{low-credibility}. Tweets without URLs or linking to unknown domains were excluded from the downstream analysis. In cases where tweets contained multiple URLs, each was treated as a separate entry, and the average domain score was assigned to the tweet.

Importantly, our ranking models operate solely on the assigned credibility of shared content, regardless of its modality. As a result, while we focus here on textual content for consistency with~\cite{deverna_identifying_2024}, the same approach naturally extends to other formats such as images, memes, or videos, provided that a credibility assessment is available. This inherent flexibility allows our methodology to adapt across media types without requiring changes to the core modeling pipeline.

\subsection{Archetype-Based User Ranking}\label{subsec:behavioral_archetypes} 

User activity within a social network offers rich signals that can be used to evaluate and compare accounts. By analyzing this activity through the lens of specific behavioral archetypes, we can define targeted methods that assign a numerical value (or \textit{score}) to each user, reflecting their role in the dissemination of misinformation. These scores are then used to generate user rankings, with each ranking capturing a different dimension of behavior based on the archetype being modeled (e.g., amplification, influence, coordination).

This section introduces a suite of archetype-based ranking methods, each designed to operationalize a specific archetypal trait. For each archetype, we formally define the corresponding ranking approach and explain the rationale behind its construction, providing a foundation for comparing how different user behaviors contribute to misinformation spread.

\subsubsection{Amplifiers Ranking}\label{subsec:amplifier_accounts_rankings}
This first section focuses on methods aimed at ranking \emph{amplifier} accounts by analyzing their resharing behavior, which reflects their role in the amplification of misinformation. Since these accounts primarily engage in re-sharing content, we first employ a method that ranks users according to the total number of misinformation posts they reshare. We then introduce a more refined approach that also accounts for the timing of these reshares, giving greater weight to users who consistently reshare earlier in the diffusion process. They are both defined as follows:

\begin{itemize}
    \item \emph{Repost Count:} This approach ranks users based on their total number of misinformation reshares during the observation period. Formally, we define the Repost Count score as:

    {\footnotesize
    \begin{equation*}\label{eq:repostcountdef}
    \text{Repost Count }(u) = \sum_{p \in \mathcal{P}_{\text{misinfo}}} R_u(p)
    \end{equation*}}

    where $\mathcal{P}_{\text{misinfo}}$ is the set of posts with credibility below our predefined threshold, and $R_u(p) = 1$ if user $u$ has reshared post $p$, otherwise $R_u(p) = 0$. While this approach offers a simple way to rank users, it overlooks the timing of their reshares—specifically, how early they amplify misinformation within its diffusion lifecycle.

    \item \emph{Early Reposter Index (EaR-Index):}  
    A more sophisticated approach ranks users based on how early they reshare low-credibility content generated by other users, considering that early re-sharing might indicate a pivotal role in information cascades. Users who systematically appear among the first to reshare viral misinformation posts receive a higher index. The EaR-index is defined as:

    {\footnotesize
    \begin{equation*}\label{eq:earindexdef}
    \text{Early Reposter Index }(u) = \frac{1}{|\mathcal{P}_u|} \sum_{p \in \mathcal{P}_u} \left( N_p - r_u(p) + 1 \right)
    \end{equation*}}
    
where $\mathcal{P}_u \subseteq \mathcal{P}_{\text{misinfo}}$ is the set of misinformation posts reshared by user $u$, $N_p$ is the total number of reshares received by post $p$, and $r_u(p)$ is the ordinal position of user $u$ in the chronologically-sorted sequence of reshares of post $p$, e.g., if $u$ was the third person to reshare $p$, then $r_u(p) = 3$. The scoring function assigns users higher values the earlier they appear in the re-sharing sequence, ensuring that those who reshare misinformation early—especially for posts that eventually receive many reshares—are ranked more prominently. The final EaR-index is computed as the average of these scores across all misinformation posts reshared by the user.
\end{itemize}

\subsubsection{Coordinated Accounts Ranking}\label{subsec:CoordinatedAccounts}

This section presents ranking methods based on coordinated activity, often modeled through behavioral similarity of user actions. The intuition behind this approach is that \emph{coordinated accounts} tend to reshare, promote, and artificially amplify the same pieces of content, reflecting a shared strategy. Following the methodology described in the literature~\cite{luceri2024unmasking,pacheco_uncovering_2021}, we construct a co-reshare network where nodes represent users, and undirected edges between users are weighted by the cosine similarity of their resharing activities. 
In line with prior work, analyzing and pruning the co-reshare similarity network enables the identification of users that play central roles in orchestrated campaigns.
To operationalize this, we propose two user ranking methods, each based on a distinct pruning strategy of the similarity network:

\begin{itemize}

    \item \emph{Coordination-Centrality:}  
    This method ranks users based on their eigenvector centrality within the co-reshare similarity network, following the rationale proposed by~\cite{luceri2024unmasking}, which leverages user centrality in similarity networks to detect \emph{coordinated accounts} involved in influence operations. A higher score indicates that user 
\textit{u} is closely connected to other central—and potentially coordinated—users, suggesting their embeddedness within organized amplification structures.

    \item \emph{Coordination-Edge Weight:}  
    This method ranks users based on the maximum weight among all edges incident to them in the similarity network, following the intuition of~\cite{pacheco_uncovering_2021}, which suggests that coordinated actors can be uncovered based on the strength of their similarity. A higher score indicates stronger coordinated behavior with at least one other user.
    
\end{itemize}

\subsubsection{Super-Spreaders Ranking}\label{subsec:super_spreaders_ranking}

This section presents a suite of approaches for ranking \emph{super-spreaders}, i.e., users who exert significant influence on social media by attracting engagement from others. Engagement can take various forms, including likes, comments, quotes, and reshares. Among these, we focus specifically on reshares, as they provide a more direct signal of content endorsement and dissemination~\cite{metaxas2015retweets}. Unlike comments or quotes, which may express disagreement, sarcasm, or criticism, reshares more reliably indicate alignment with the content and contribute directly to its amplification. 

The following methods are used to assign scores to users and subsequently rank them based on their ability to spread content and elicit engagement from others. We categorize these methods into two groups: state-of-the-art approaches and our proposed methods:

\begin{itemize}
    \item \emph{State-of-the-art methods}, as proposed by DeVerna et al.,~\cite{deverna_identifying_2024}, use the volume of retweets received as a proxy for user influence. Specifically, they introduce two approaches, namely:

    \begin{itemize}
        \setlength\itemsep{1em}
        \item \emph{Influence Score:} This method calculates the total number of reshares received by an account’s low-credibility posts, providing a direct measure of the user's overall reach in spreading such content:

        {\footnotesize
        \begin{equation*}
        \text{Influence Score }(u) = \sum_{p \in \mathcal{P}_u} \text{Reposts}_p
        \end{equation*}}
        
        where \(\mathcal{P}_u\) is the set of low-credibility posts created by user \(u\), and \(\text{Reposts}_p\) is the number of reshares gathered by post \(p\). 

        \item \emph{H-Index:} This method treats posts as publications and reshares as citations. The H-Index, \(h\), for a user \(u\) is defined as the largest number of posts \(h\) that have received at least \(h\) reshares each. Formally:

        {\footnotesize
        \begin{equation*}
        \text{H-Index }(u) = \max \{ h \mid u \text{ has } h \text{ posts} \in \mathcal{P}_u \text { with reshare count} \geq h\}
        \end{equation*}}

        This method ensures that a user's influence is not determined by a single viral post but instead reflects a consistent ability to generate engagement.
    \end{itemize}

However, these two methods compute user scores within a fixed, and often arbitrarily chosen, observation period, meaning that the resulting rankings may be influenced by both the length of the window and the frequency of user activity. This can introduce inconsistencies when comparing users with different posting patterns. Notably, the H-index offers a more stable alternative, as it is less sensitive to temporal constraints: users must maintain consistent engagement over time to increase their H-index, making it more robust to variability.

    \item \emph{Our Proposed Time-Aware Methods:} Since user activity varies over time, we partition the observation window into non-overlapping, contiguous time intervals and compute scores within each slot. To aggregate these values, we apply an exponential moving average, which smooths the scores across time while giving more weight to recent behavior. This approach balances historical influence with recent activity, capturing long-term trends while reducing the impact of short-term fluctuations. Based on this framework, we define two time-aware variants of the state-of-the-art methods:

    \begin{itemize}
        \item \emph{Time-Aware Influence Score,}
        which applies the Influence Score method for each time slot. These per-slot scores are then smoothed over time using an exponential moving average, resulting in a final score that reflects both historical and recent influence in terms of content virality. The Time-Aware Influence (TAI) Score is defined as follows:

        {\footnotesize
        \begin{equation*}\label{eq:tais_def}
        \begin{cases}
        \text{TAI-Score }(u)_t = \alpha \cdot \text{TAI-Score }(u)_{t-1} + (1 - \alpha) \cdot \text{Influence Score }(u)_t
        \\
        \text{TAI-Score }(u)_0 = \text{Influence Score }(u)_0
        \end{cases}
        \end{equation*}}

        where \(\alpha\) is the smoothing factor, and \(\text{Influence Score }(u)_t\) is the Influence Score for user \(u\)  at time slot \(t\) spanning \(\delta\) days. Intuitively, a higher \(\alpha\) assigns more weight to past influence, while a lower \(\alpha\) emphasizes recent activity.\\

        \item \emph{Time-Aware Social H-Index:} Similar to the TAI-Score, this method applies an exponential moving average to the H-index computed within each time slot. This produces a smoothed score that captures a user’s sustained ability to generate widely reshared content over time, balancing historical consistency and recent activity. The Time-Aware Social Hirsch (TASH) Index is defined as follows:

        {\footnotesize
        \begin{equation*}\label{eq:tashidef}
        \begin{cases}
        \text{TASH-Index }(u)_{t} = \alpha \cdot \text{TASH-Index }(u)_{t-1} + (1 - \alpha) \cdot \text{H-Index }(u)_t
        \\
        \text{TASH-Index }(u)_{0} = \text{H-Index }(u)_0
        \end{cases}
        \end{equation*}}
        
        where $\text{H-Index }(u)_t$ is the user \(u\)'s H-Index as proposed by~\cite{deverna_identifying_2024}, computed at time slot \(t\) spanning \(\delta\) days, while \(\alpha\) controls the trade-off between long-term and short-term influence.
    \end{itemize}
\end{itemize}

In conclusion, both time-aware methods require tuning two parameters: \(\delta\), the size of the time slots \(t\) in days, and \(\alpha\), the smoothing factor. When \(\alpha = 1\), the TASH-Index remains constant, equal to the H-index from the first observed time slot, effectively ignoring any updates over time. As \(\alpha\) increases, it adjusts the balance between historical estimates and recent influence. When \(\alpha = 0\) instead, we consider only the H-index for the user in the most recent time slot of size \(\delta\). To determine the best values for \(\alpha\) and \(\delta\), we conduct a grid search using the \emph{normalized Discounted Cumulative Gain} metric~\cite{kalervo2002ndcg} to evaluate different ranking performances. Further details on the search space are provided in~\ref{appendix:optimization}.

\subsection{Machine Learning-based Ranking}\label{subsec:ml_ranking}

The ranking methods introduced thus far rely on distinct behavioral traits, each designed to capture a specific archetypal aspect of user behavior. Building on this foundation, we explore whether combining these traits can improve ranking accuracy. Moreover, by integrating multiple archetypal traits into a unified model, we aim to interpret the model’s decisions and gain deeper insights into the key factors driving misinformation spread.

Ranking items is a well-established task in machine learning, commonly referred to as \textit{learning to rank}~\cite{Liu2009LearningTR}, and it aligns naturally with our objective. 

To comprehensively evaluate the effectiveness of different approaches, we frame the user ranking problem as a regression task—also known as pointwise ranking~\cite{li_hang_learning_to_rank, Melnikov_Hüllermeier_Kaimann_Frick_Gupta_2016}. By modeling the ranking as a regression problem, we aim to predict each user's future contribution to misinformation based on their historical behavior and extracted features. This formulation provides a systematic framework for assessing user influence and enables meaningful comparisons with alternative ranking methods in our evaluation.

The \emph{target variable} for this regression task quantifies a user’s \emph{strength} in the misinformation reshare network. Specifically, it is defined as the total number of reshares involving low-credibility content in which the user is engaged—either as the original author whose posts are amplified by others, or as a user who actively reshares such content. This metric captures both the user's influence and their activity level in the propagation of misinformation, offering a comprehensive indicator of their role in its diffusion.

Each user is represented by a feature vector comprising behavioral indicators drawn from the archetypes previously introduced, including:

\begin{itemize}
    \item Repost Count \emph{(Repost Count)}%
    \item Early Reposter Index \emph{(EaR-Index)}%
    \item Coordination-Centrality \emph{(Coord-Centrality)}%
    \item Coordination-Edge Weight \emph{(Coord-EdgeWeight)}%
    \item Time-Aware Influence Score \emph{(TAI-Score)}%
    \item Time-Aware Social H-Index \emph{(TASH-Index)}%
\end{itemize}

We train and test several off-the-shelf ML models, with Random Forest resulting in the most accurate one. Finally, we apply SHAP (SHapley Additive exPlanations) to interpret the model’s predictions by identifying the features that most significantly influence the output. This analysis not only highlights the behavioral traits driving misinformation spread but also helps surface the users who play the most critical roles in its diffusion.

\section{Evaluation}\label{sec:evaluation}

In this section, we first describe the data sources used in our study, detailing their structure and relevance to the analysis. We then present the evaluation methodology and define the metrics employed to assess the performance of the ranking models introduced in the previous section.

\subsection{Datasets}\label{subsection:datasets}

This study relies on two datasets, both collected from Twitter during a period of heightened misinformation diffusion related to the COVID-19 pandemic. To evaluate the consistency of our findings across different contexts, we selected datasets focused on the same topic but covering distinct geographic areas: one limited to a single country (Italy), and the other reflecting the global conversation—albeit with a notable U.S. bias, given the platform’s user base. These datasets are described as follows:

\begin{itemize}
    \item \emph{VaccinItaly}: This dataset, collected by Pierri et al.,~\cite{pierri2021vaccinitalymonitoringitalianconversations}, contains COVID-19-related tweets from Italian-speaking accounts, focusing on discussions related to vaccines and public health from December 20, 2020, to October 22, 2021 (306 days). Considering only reshares that can be scored—i.e., those containing one or more URLs—the final dataset includes 371,586 retweets from 54,916 distinct original posts and comprising 51,962 unique users. 

    \item \emph{COVID-19 Multilanguage}: This larger multilingual dataset of tweets was collected during the pandemic~\cite{giovanni2022vaccineu} and spans 372 days, from November 1, 2020, to November 8, 2021. The dataset includes 1,118,697 reshares from 103,250 distinct original posts and comprises 129,507 unique users.

\end{itemize}

\begin{figure}[t]
    \centering
    \begin{subfigure}[b]{0.48\textwidth}
        \centering
        \includegraphics[width=\linewidth]{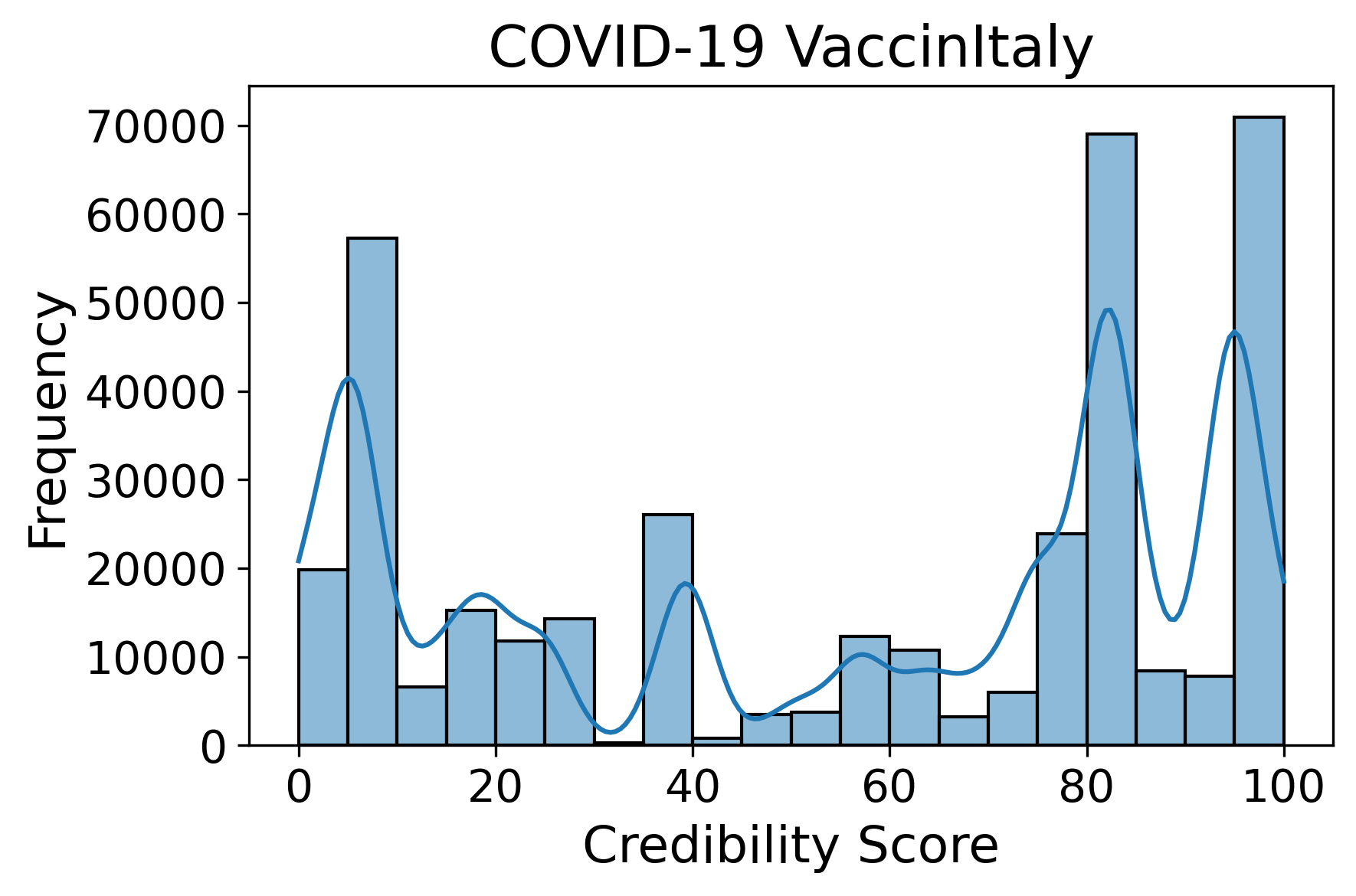}
        \label{fig:vaccinitaly_cred_distr}
    \end{subfigure}
    \hspace{0.2cm}
    \begin{subfigure}[b]{0.48\textwidth}
        \centering
        \includegraphics[width=\linewidth]{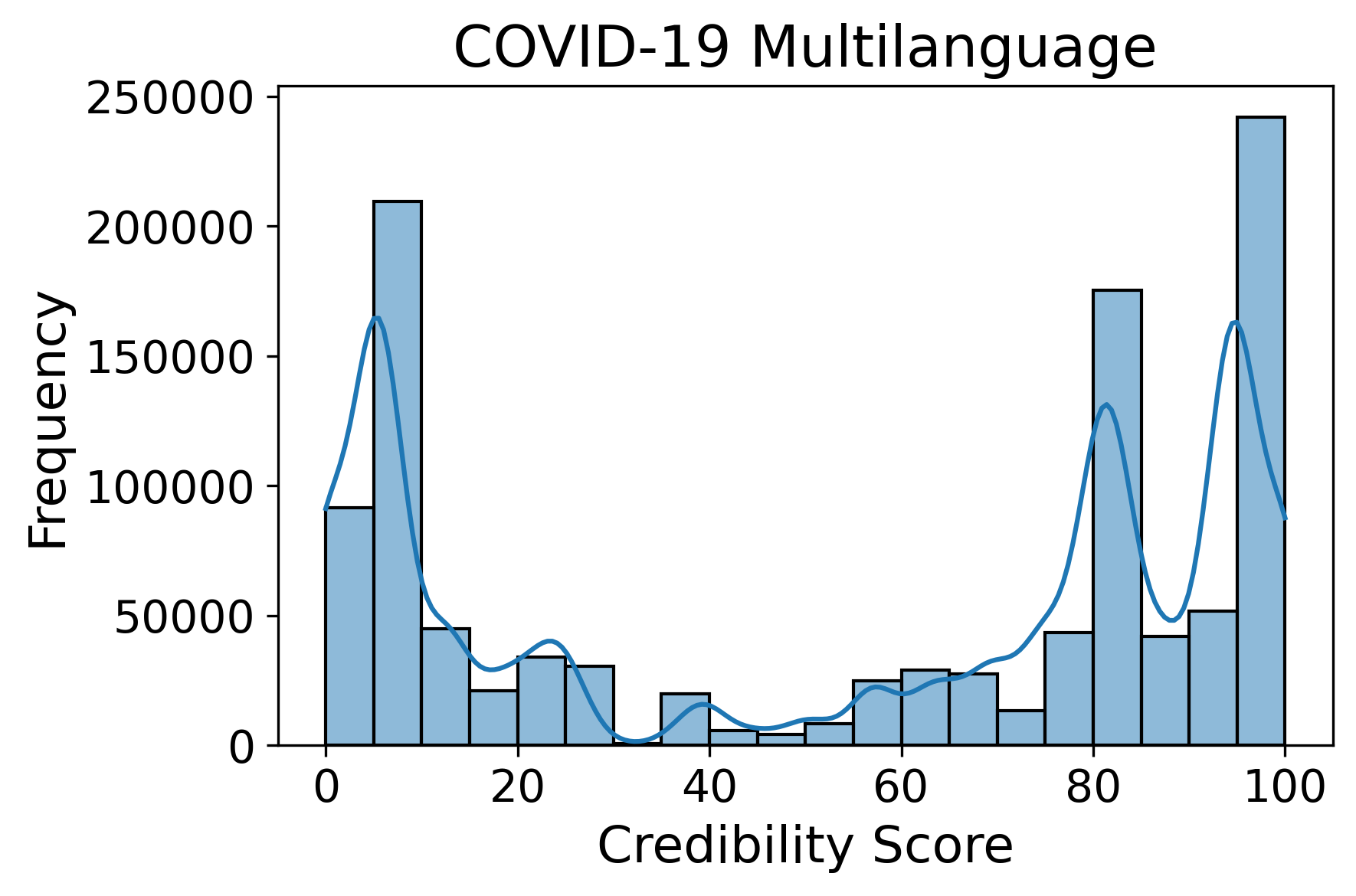}
        \label{fig:multilang_cred_distr}
    \end{subfigure}
    \caption{\emph{Credibility distribution for labeled retweets for each dataset under scrutiny}: VaccinItaly (left) and COVID-19 Multilanguage (right). Credibility score is assigned to each tweet based on NewsGuard's ratings, as described in Section~\ref{subsec:credibility_assessment}}
    \label{fig:cred_distrib}
\end{figure}

Figure~\ref{fig:cred_distrib} presents the distribution of credibility scores associated with shared URLs in the two datasets. Both plots capture the credibility landscape of online content circulated during the COVID-19 pandemic, reflecting different linguistic and geographic dynamics.
In the VaccinItaly dataset (left panel), the distribution is bimodal, with clear peaks in the very low credibility range (0–10) and the very high range (90–100). However, the low-credibility peak is less sharp and more spread across the 0–40 range than in the Multilanguage dataset. The presence of a secondary concentration around credibility scores in the 70–80 range suggests a more varied credibility spectrum within the Italian conversation, rather than strict polarization. Still, the data reveal a tendency for users to share content from both low- and high-credibility sources, with relatively less activity in the mid-range (40–70). 
In the Multilanguage dataset (right panel), the distribution is more sharply skewed, with a dominant peak in the lowest credibility bracket (0–10) and a strong resurgence at the top end (90–100). Unlike the Italian dataset, there is less evidence of mid-credibility sharing, and the low-credibility activity is more concentrated around the extreme values. This indicates a more polarized global discourse, particularly driven by content from highly questionable or highly reputable domains, with fewer intermediate sources in between.

By leveraging two datasets that differ in scale, geographic focus, and linguistic diversity, we ensure that our evaluation captures variability in user behavior and information diffusion dynamics across distinct ecosystems.

\paragraph{Data splitting: Observation vs. Evaluation Periods} We adopt a temporal split strategy to ensure that no information from the future leaks into model training. Specifically, the \emph{observation period} is used as the \emph{training set}—that is, all data the model can access—while the \emph{evaluation period} serves as the \emph{test set}, following a standard 80--20 proportion. This split is applied \emph{sample-wise} in chronological order, selecting the first 80\% of resharing events in their natural temporal sequence. Such a split preserves temporal consistency and respects causality, ensuring that the model never uses future events to predict past behavior. Within the training portion, we apply a sliding-window approach to construct temporally valid validation sets, following the cross-validation scheme described in~\ref{appendix_ml_tuning}.

\subsection{Network Dismantling}\label{subsec:network_dismantling}  

To evaluate the accuracy of distinct ranking strategies, we apply a network dismantling approach, ensuring a fair comparative analysis with previous work~\cite{deverna_identifying_2024}. We construct a reshare network by modeling it as a weighted, directed graph where nodes represent users and directed edges represent reshare activities. The weight of each edge corresponds to the total number of distinct pieces of content reshared between two users. For instance, if an edge exists from user \(A\) to user \(B\) with weight 3, this indicates that \(B\) reshared three distinct pieces of content originally created by \(A\).  

To ensure consistency across platforms, we count reshares at the level of unique content. This decision accounts for differences in how platforms handle resharing: some, like X (formerly Twitter), prevent users from resharing the same post multiple times, while others, such as Facebook, allow repeated sharing of the same content. As a result, the edge
 \(A \xrightarrow[]{3} B\)  indicates that user 
\(B\) has reshared exactly three distinct pieces of content originally posted by user 
\(A\), regardless of how many times each piece was shared. From this network, we construct a \emph{low-credibility reshare network} by retaining only content with credibility scores below a predefined threshold (39, based on NewsGuard guidelines). This filtered network captures the dissemination patterns of misinformation among users over the analyzed time period.

The network dismantling process involves iteratively removing users from the low-credibility reshare network, based on a given ranking. Removing a node eliminates all its incident edges, effectively preventing the associated misinformation reshares. We quantify the impact of removing a specific user by summing the weights of all incident edges (both incoming and outgoing) and normalizing this value as a fraction of the total network weight.

While this method does not account for potential cascading effects—such as whether removing one user influences others' behavior—it serves as a practical proxy for estimating the impact of targeted user moderation. Specifically, it quantifies the share of misinformation that would no longer circulate if certain users were deplatformed.

\subsection{Evaluation Metrics}\label{subsec:evaluation_metrics}

\begin{figure}[t]
    \centering
    \includegraphics[width=\textwidth]{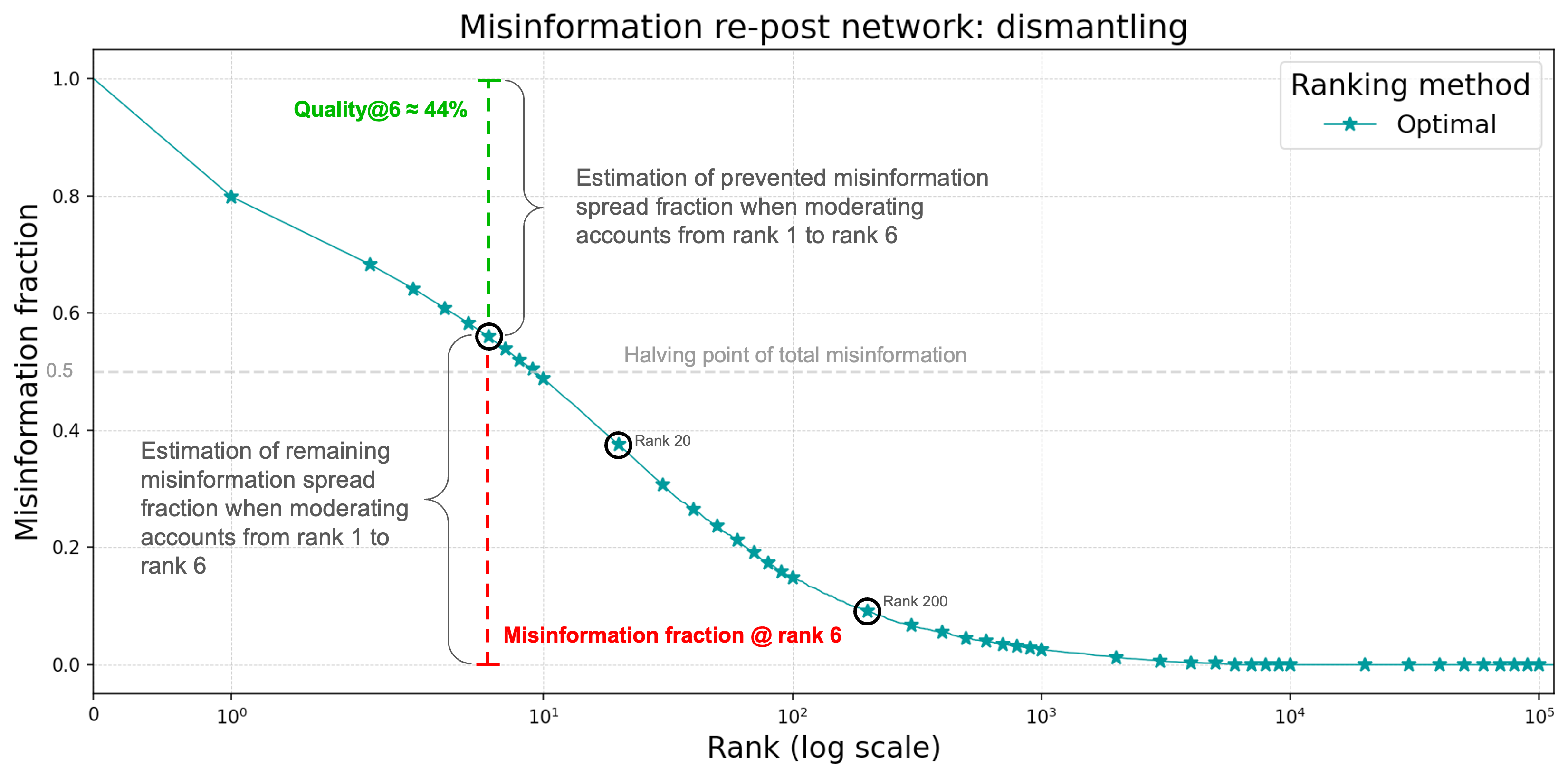}
    \caption{\emph{Illustration of the network dismantling process.} The figure illustrates how the removal of highly ranked users impacts the overall volume of misinformation. Each point on the curve corresponds to one user removed from the network, in descending order according to their ranking. The observed trend reveals that a small subset of users is responsible for a disproportionately large share of misinformation diffusion.}
    \label{fig:dismantling_explainer}
\end{figure}

We assess the performance of the different ranking models using two complementary metrics: \emph{Quality@k}, which we derived from the network dismantling procedure introduced in prior work~\cite{deverna_identifying_2024} (and illustrated in Figure~\ref{fig:dismantling_explainer}), and \emph{nDCG@k} (normalized Discounted Cumulative Gain)~\cite{kalervo2002ndcg}. Both metrics are computed by evaluating model-generated user rankings against the low-credibility reshare network constructed from the test set, representing unseen misinformation dissemination patterns.
The dismantling procedure shown in Figure~\ref{fig:dismantling_explainer} simulates the sequential removal of top-ranked users from the reshare network and measures the resulting drop in misinformation volume.

The curve represents the optimal dismantling, obtained by ranking users in descending order according to ground truth from the evaluation (test) set—specifically, their total node strength in the low-credibility reshare network. Node strength is defined as the sum of a user's incoming and outgoing reshare connections, reflecting both their role in amplifying others and being amplified. Each point on the curve corresponds to the removal of a user according to this ranking, highlighting the cumulative reduction in misinformation volume as top-ranked users are sequentially removed.

\emph{Quality@k} estimates the proportion of misinformation that would be removed by moderating the top-$k$ ranked users in the test set network. For example, \emph{Quality@5} represents the share of low-credibility content eliminated by suppressing the top 5 users in the ranking. This metric provides an intuitive measure of ranking effectiveness for practical moderation scenarios.\footnote{This is a conservative estimate of the actual impact of moderation, as it assumes the complete removal of content that would have been generated by moderated users.} \emph{nDCG@k}~\cite{kalervo2002ndcg} is a standard information retrieval metric that evaluates ranked output quality by considering both item relevance and position, assigning higher scores when relevant items appear earlier in the ranking. In our context, an item correspond to a user and each user's relevance is determined by their volume of misinformation dissemination during the test period. The truncated version focuses only on the top-$k$ users in both predicted and ground-truth rankings, ensuring comparability with \emph{Quality@k}.

These metrics provide complementary perspectives: \emph{Quality@k} captures the potential moderation benefit of each method when applied to future misinformation networks, while \emph{nDCG@k} reflects the consistency between predicted and actual user influence in misinformation spread during the test period.

\paragraph{Evaluation setup} Our models produce user rankings based on training data, and we measure their effectiveness by applying these rankings to dismantle the misinformation reshare network observed in the test period. This evaluation design reflects the practical scenario where moderation decisions must be made based on historical user behavior to prevent and mitigate future misinformation spread.

Unlike standard recommendation tasks with static item sets, our scenario involves a dynamic user base, which can lead to discrepancies between training and test sets. We address this by distinguishing two user categories:

\begin{itemize}
    \item \textit{Users present in training but absent in test} are assumed to contribute no future misinformation. They are retained in the predicted ranking and assigned a ground-truth relevance score of zero.
    \item \textit{Users appearing only in the test set }are considered unrecoverable by the model, as they were unobserved during training. These users are excluded from evaluation.
\end{itemize}

This design ensures that predicted and ground-truth rankings are aligned, matching in length and referring to the same set of users. While this setup introduces a systematic downward bias by omitting unseen future users, it reflects a realistic modeling constraint: a model can only rank users it has previously observed.

\section{Results}\label{sec:results_analysis}

This section presents the performance of each model using the metrics previously outlined.

Special emphasis is placed on explainability, utilizing the SHAP technique to offer insights into model decisions and feature contributions to the model outcome. In addition, we examine the influence of data scarcity on the performance of the best-performing methods. This analysis sheds light on the robustness and adaptability of the models under varying data availability scenarios.

\subsection{Ranking Performance}\label{subsec:performances}

This subsection presents a comparative analysis of the best-performing ranking methods for each user archetype, using the dismantling evaluation framework introduced in Section~\ref{subsec:network_dismantling}. A more extensive evaluation involving all the ranking methods is provided in~\ref{appendix_rankings}.
Figure~\ref{fig:dismantling_overall} illustrates the dismantling trajectories obtained by progressively removing users from the VaccinItaly dataset according to different ranking strategies. Each curve tracks the cumulative fraction of misinformation removed as more users are blocked, simulating a moderation intervention.

Figure~\ref{fig:dismantling_overall} includes the best method per archetype, as well as an optimal dismantling track (star-marked curve) that serves as our ground truth. The objective is to evaluate the effectiveness of each method in mitigating misinformation and, in turn, to identify which user archetypes are most relevant or most reliably detected within the analyzed framework. In practical terms, we are most interested in the early portion of each curve: the left-hand side indicates how much misinformation is mitigated by removing only the top-ranked users. Conversely, the tail of the curve reflects diminishing returns and is less relevant for real-world scenarios. These removals assume complete blocking of users, without modeling the dynamic effects of moderation on user behavior or network structure. Nevertheless, they provide useful insights into the relative effectiveness of different approaches.

\begin{figure}[t]
    \centering
    \includegraphics[width=\textwidth]{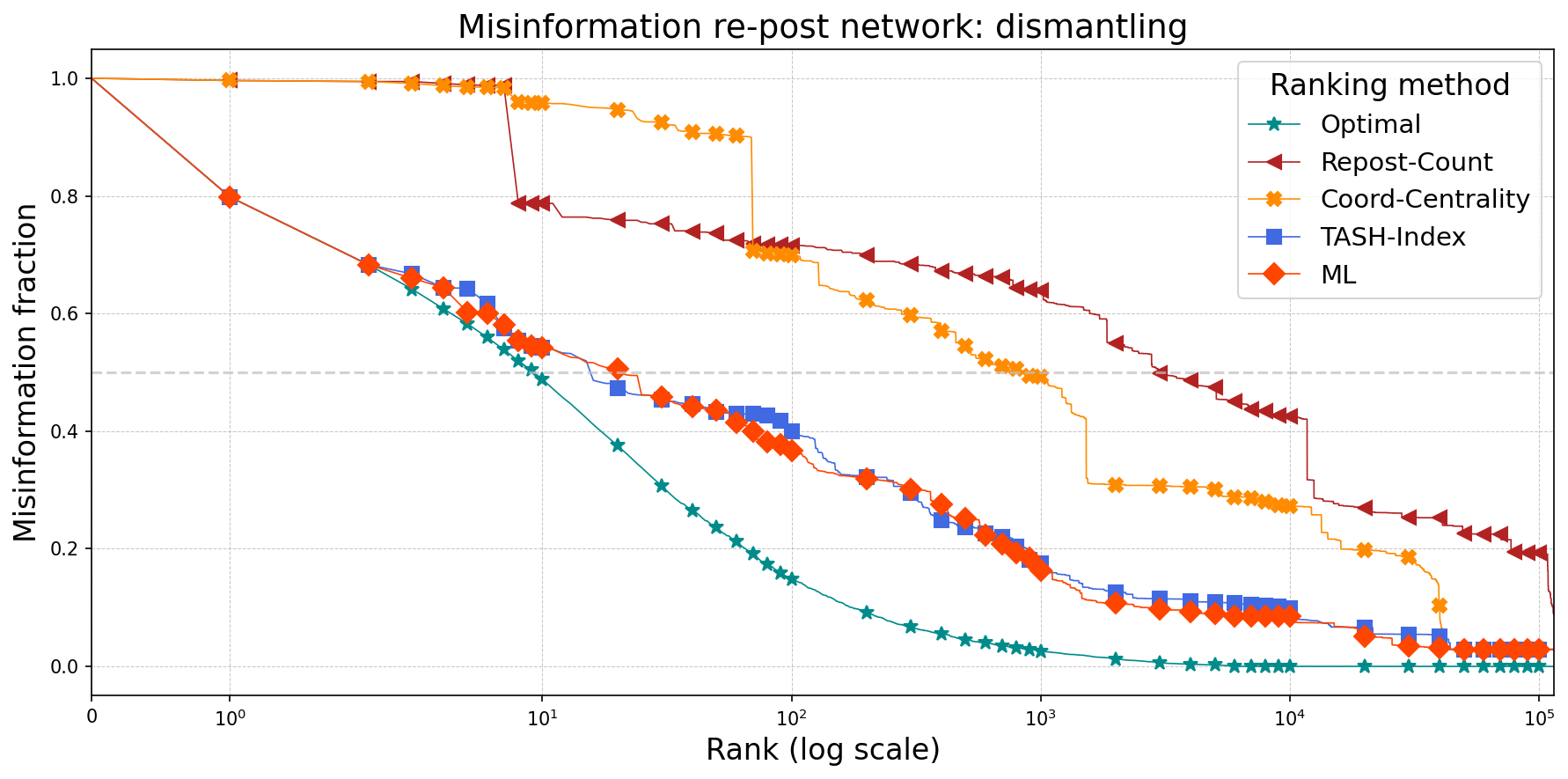}
    \caption{\emph{Performance of the best performing ranking methods for each archetype.} Each marker represents a user removed from the misinformation reshare network of the VaccinItaly dataset. The x-axis represents the user’s rank, while the y-axis indicates the estimated reduction in misinformation achieved by removing that user.}
    \label{fig:dismantling_overall}
\end{figure}

A key finding from this analysis is that the ML-based method and TASH (a method grounded in the \emph{super-spreader} archetype) achieve performance comparable to the optimal ranking, particularly for the top-ranked users. Moreover, both methods consistently outperform alternative strategies in identifying the most influential actors in the spread of misinformation.

This can be attributed to how these rankings align with the structure of the reshare network: the most impactful nodes in this network are those with high \textit{node strength}, defined as the sum of incoming and outgoing edge weights, which in turn indicate produced and received retweets. Nodes with high strength, therefore, indicate users who are highly effective at spreading content and receive substantial engagement from others. Therefore, ranking strategies that prioritize such properties, such as the volume of reshares received, naturally result in better performance.

\begin{table}[t]
\centering
\resizebox{\textwidth}{!}{
\begin{tabular}{lcc|cc|cc|cc|cc}
\toprule
 & \multicolumn{2}{c|}{\textbf{ML}} & \multicolumn{2}{c|}{\textbf{TASH-Index}} & \multicolumn{2}{c|}{\textbf{H-Index}} & \multicolumn{2}{c|}{\textbf{TAI-Score}} & \multicolumn{2}{c}{\textbf{Influence Score}} \\
\cmidrule(lr){2-3} \cmidrule(lr){4-5} \cmidrule(lr){6-7} \cmidrule(lr){8-9} \cmidrule(lr){10-11}
@k  & nDCG  & Quality  & nDCG  & Quality  & nDCG  & Quality  & nDCG  & Quality  & nDCG  & Quality  \\
\midrule
3    & \textbf{0.963}  & \textbf{34.0\%}  & \underline{0.949}  & \underline{33.3\%}  & 0.892  & 32.6\%  & 0.938  & 32.6\%  & 0.892  & 32.6\%  \\
5    & \textbf{0.967}  & \textbf{39.8\%}  & \underline{0.912}  & \underline{35.7\%}  & 0.858  & 35.0\%  & 0.901  & 35.0\%  & 0.865  & 35.7\%  \\
10   & \textbf{0.939}  & \textbf{45.8\%}  & \underline{0.931}  & \textbf{45.8\%}  & 0.832  & 40.4\%  & 0.873  & 40.3\%  & 0.824  & 40.4\%  \\
20   & 0.890  & 49.0\%  & \textbf{0.906}  & \textbf{52.8\%}  & 0.790  & 43.7\%  & \underline{0.848}  & \underline{49.2\%}  & 0.783  & 43.2\%  \\
100  & \textbf{0.848}  & 62.3\%  & \underline{0.832}  & 60.0\%  & 0.800  & \underline{66.7\%}  & 0.829  & 65.8\%  & 0.793  & \textbf{66.9\%}  \\
1000 & \textbf{0.841}  & 81.8\%  & \underline{0.782}  & 82.4\%  & 0.742  & 81.7\%  & 0.774  & \textbf{83.9\%}  & 0.735  & 82.7\%  \\
\midrule
\textbf{Overall} & \textbf{0.904}  & 97.1\%  & \underline{0.868}  & 97.1\%  & 0.829  & 97.1\%  & 0.855  & 97.1\%  & 0.821  & 97.1\%  \\
\bottomrule
\end{tabular}
}
\caption{\emph{Comparison of methods using nDCG and Quality scores at different @k levels on the VaccinItaly Dataset.} 
The maximum Quality score remains identical across methods when all the nodes are removed from the misinformation reshare network, except those that appear after the training phase; consequently, 2.9\% of misinformation cannot be mitigated due to the emergence of new users in the test set. 
\textbf{Bold} values denote the best results, \underline{underlined} values indicate the second best.}

\label{tab:ml_vs_ss_vaccinitialy}
\end{table}
 
Table~\ref{tab:ml_vs_ss_vaccinitialy} presents a comparative, quantitative evaluation of the methods presented in Figure~\ref{fig:dismantling_overall}, using the two key metrics introduced before: nDCG and Quality score. 
Overall, the ML model exhibits the best ranking performance, achieving the highest nDCG score (0.904) and outperforming all other methods, particularly at small and moderate intervention levels (k = 3 to 100). The TASH-Index, despite its relative simplicity, closely follows it in terms of ranking performance, showing strong results particularly in top rankings. Quality scores at these levels also indicate that ML may mitigate the largest share of misinformation with fewer interventions, followed by TASH-Index. This aligns with previous findings that show that a limited number of users (small $k$) is responsible for a large amount of misinformation~\cite{Nogara2022TheDD,baribi2024supersharers}. As the value of k increases, the TAI-Score emerges as particularly effective, attaining the highest Quality score (83.9\%) at k = 1000, suggesting its potential strength for broad-scale interventions. 
In contrast, the H-Index and Influence Score consistently underperform across both evaluation metrics. This further underscores the importance of incorporating temporal dynamics. Such temporal awareness proves particularly effective in detecting users who contribute to the rapid spread of harmful content, making time-aware methods substantially more informative than static ones.  

Overall, our analysis highlights a clear trade-off between model complexity and effectiveness: while ML offers superior performance, heuristic-based methods like TASH-Index can approximate its impact at lower computational costs.
The TAI-Score, however, demonstrates value when wide-scale intervention is possible or necessary. 

The results from the COVID-19 Multilanguage dataset closely mirror those observed in the VaccinItaly dataset (see Table~\ref{tab:ml_ss_multilanguage} in the Appendix), reinforcing the robustness and generalizability of our findings across different linguistic and geographic contexts.

\subsection{Machine Learning Explanation}\label{subsec:ml_explanations}

The machine learning component of our framework is designed to rigorously evaluate the effectiveness of combining multiple behavioral archetype features to characterize online user behavior—an area that remains underexplored in prior literature. By leveraging interpretable and well-established modeling techniques, our approach not only facilitates accurate ranking of users by their misinformation spreading potential, but also provides explanatory insight into the relative influence of different behavioral traits. This enables a deeper understanding of the mechanisms behind misinformation dissemination and offers actionable directions for intervention strategies.

To better understand and interpret the predictions of the machine learning model, we adopt SHAP~\cite{lundberg2017unifiedapproachinterpretingmodel}, a widely used framework that quantifies the contribution of each input feature to the model's output. This enhances both interpretability and transparency, allowing us to uncover which behavioral archetypes most strongly drive the ranking outcomes.

While the TASH-Index is inherently interpretable, machine learning models are typically opaque. SHAP mitigates this limitation by decomposing each prediction into additive contributions from individual features. In our setting, the model is intentionally constrained to use only archetypal features. This constraint ensures that the model’s decisions are grounded in the predefined behavioral profiles.
Crucially, this setup allows for a meaningful comparison between the performance and interpretability of the TASH-Index and the ML model, offering complementary insights into the behavioral signals that characterize key misinformation spreaders.

\begin{figure}[t!]
    \centering
    \includegraphics[width=\textwidth]{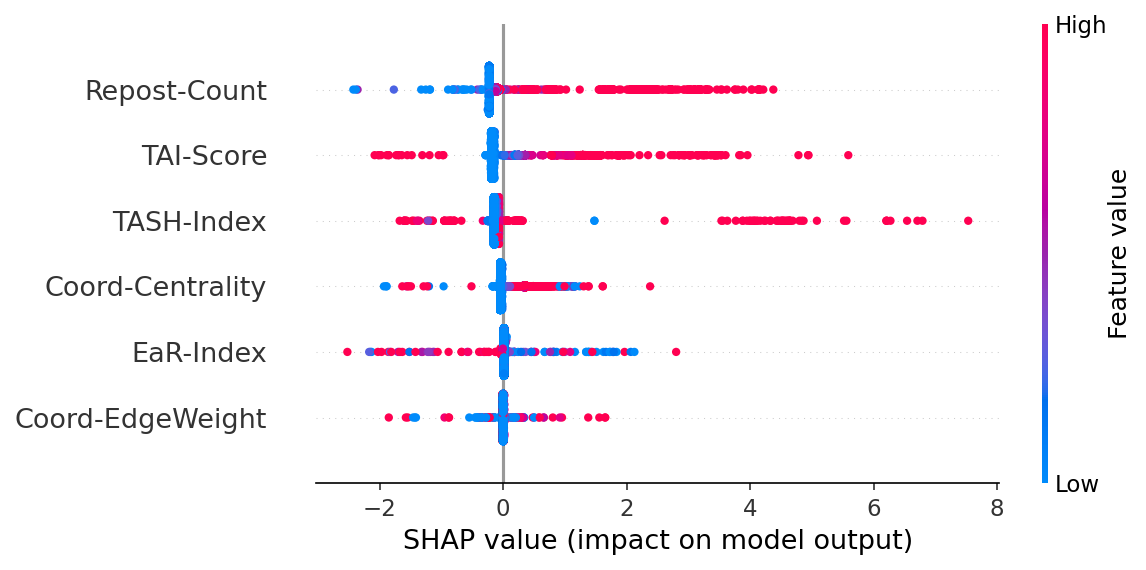}
    \caption{\emph{SHAP summary plot:} The plot displays the impact of each feature on the machine learning model’s predictions. Each dot represents a SHAP value for a particular user-feature instance, with color indicating the actual feature value (from low in blue to high in red). SHAP values quantify the direction and magnitude of a feature's contribution to the model's output. Values are shown on a logarithmic scale to enhance visibility and interpretability.}
    \label{fig:shap_summary}
\end{figure}

Figure~\ref{fig:shap_summary} shows the SHAP summary plot for the ML regression model. Each point represents a user, with the color indicating the feature value (red for high, blue for low), and the position along the x-axis representing the impact of that feature on the user's predicted ranking. The most influential feature is the \emph{Repost Count}, associated with the \emph{amplifier} archetype, which quantifies the volume of low-credibility content a user reshared. 

The color gradient in the SHAP summary plot adds interpretability by linking feature values to their impact on model predictions. In the case of the Repost Count feature, the plot reveals a robust positive correlation between a user’s frequency of amplifying low-credibility content and their predicted ranking. That is, the model consistently learns that frequent resharing behavior is a strong signal to detect pivotal actors in misinformation dissemination. 

Interestingly, although the Repost Count feature is not directly linked to the super-spreader archetype, it emerges as the most important feature in the prediction model. This finding is somewhat unexpected, as one might assume that features associated with influence---such as those defining the super-spreader archetype---would exert the strongest effect. However, this finding underscores the critical role of amplifiers in the information ecosystem and shows that resharing alone can serve as a dominant behavioral indicator for identifying relevant actors in the spread of misinformation.

The next two most influential features identified by the model are the \textit{TAI-Score} and the \textit{TASH-Index}, both linked to the \textit{super-spreader} archetype. In both cases, higher values correspond to higher predicted rankings, indicating that the model effectively identifies users who generate original content that triggers wide reshare cascades and play a central role in initiating the spread of misinformation. This reinforces the role of super-spreaders as central actors in the misinformation ecosystem. Other features, such as Coordination-Centrality, EaR-Index, and Coordination-EdgeWeight, exhibit lower but still meaningful contributions to the model outcome.

Overall, the SHAP summary analysis confirms that the model ranks users based on nuanced combinations of archetypal features. While it assigns substantial importance to behaviors typical of the \emph{super-spreader} archetype---such as high TAI-Score and TASH-Index---it consistently prioritizes features linked to the \emph{amplifier} archetype, with the Repost Count emerging as the dominant predictor. This is particularly interesting, as \emph{amplifier}-like behavior alone may not fully characterize high-risk users, yet when combined with other signals, it plays a critical role in the model’s decision-making. By integrating multiple archetypal dimensions, the model surpasses simpler strategies based on a single archetype.

\begin{figure}[t!]
    \centering
    \hspace{-0.0cm}
    \makebox[\textwidth]{
    \scalebox{1.05}{
        \begin{minipage}{1.0\textwidth}
            \centering
            \begin{subfigure}[b]{0.45\textwidth}
                \centering
                \includegraphics[width=\linewidth]{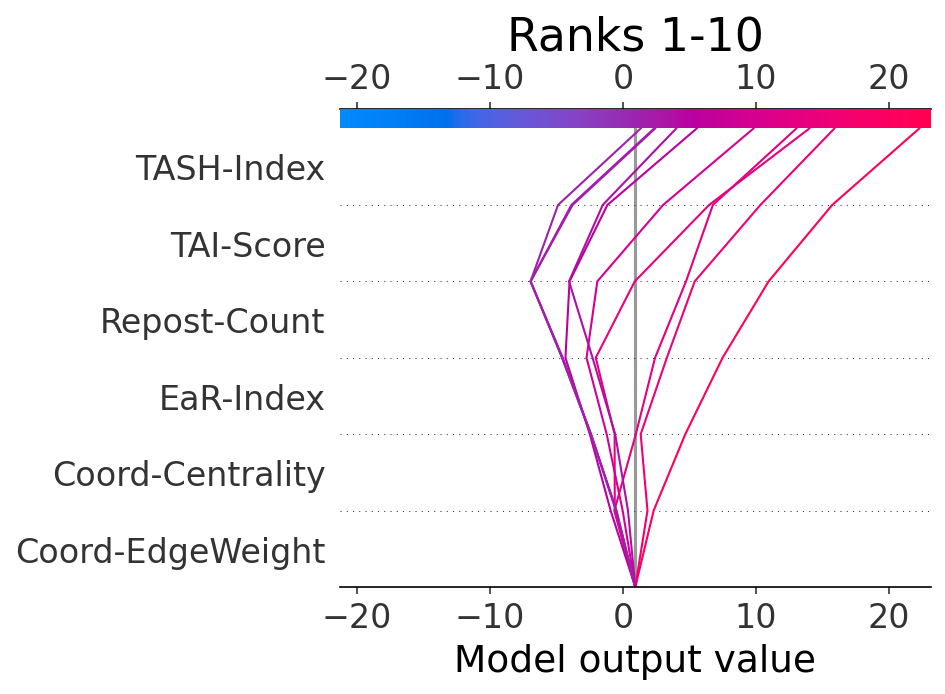}
                \label{fig:sub1}
            \end{subfigure}
            \hspace{0.1cm}
            \begin{subfigure}[b]{0.45\textwidth}
                \centering
                \includegraphics[width=\linewidth]{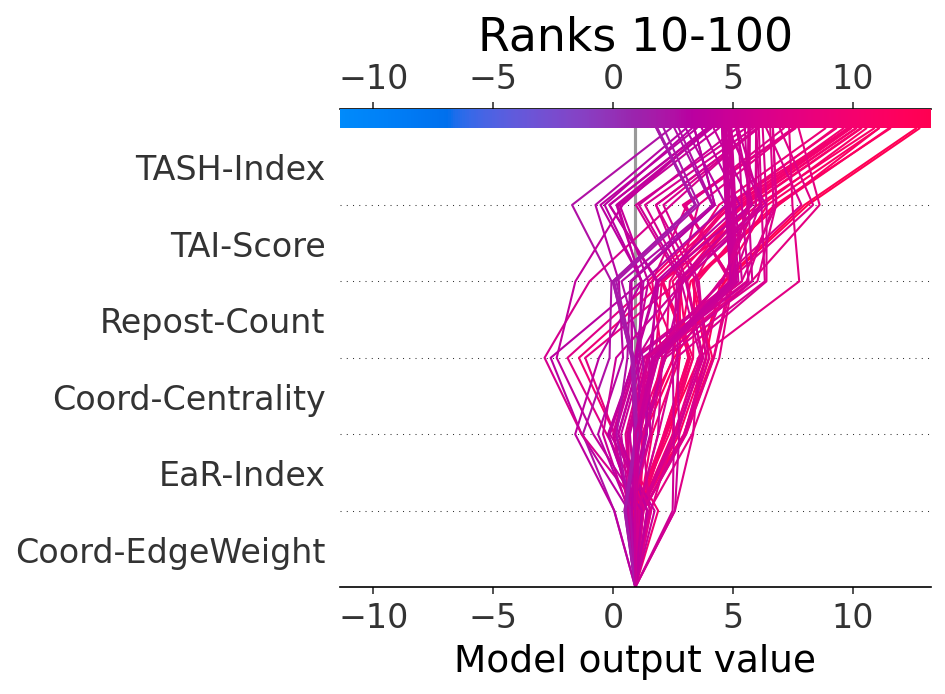}
                \label{fig:sub2}
            \end{subfigure}
            
            \vspace{0.1cm}
            
            \begin{subfigure}[b]{0.45\textwidth}
                \centering
                \includegraphics[width=\linewidth]{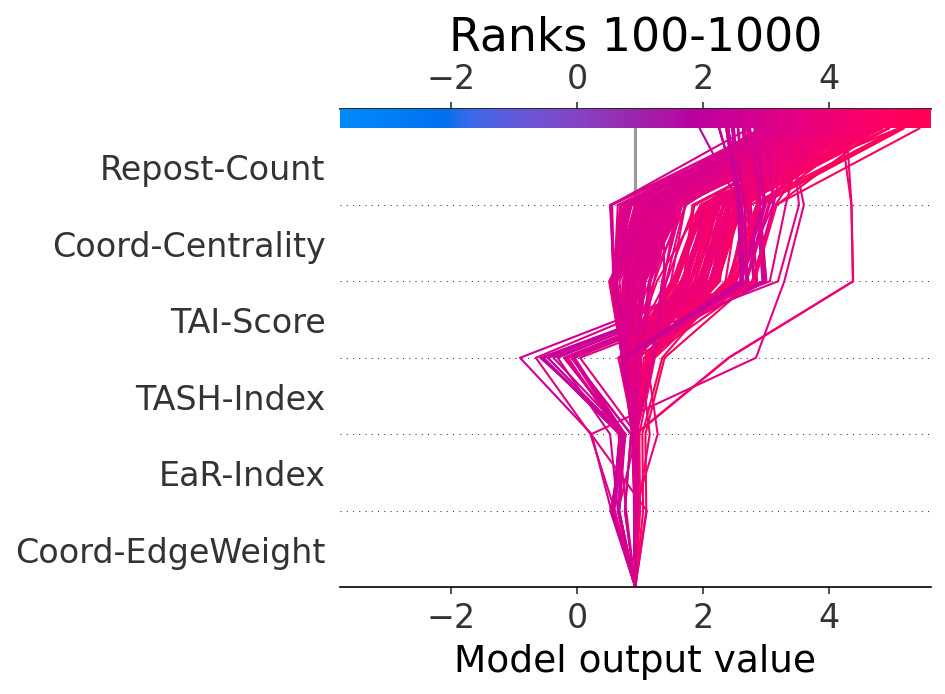}
                \label{fig:sub3}
            \end{subfigure}
            \hspace{0.1cm}
            \begin{subfigure}[b]{0.45\textwidth}
                \centering
                \includegraphics[width=\linewidth]{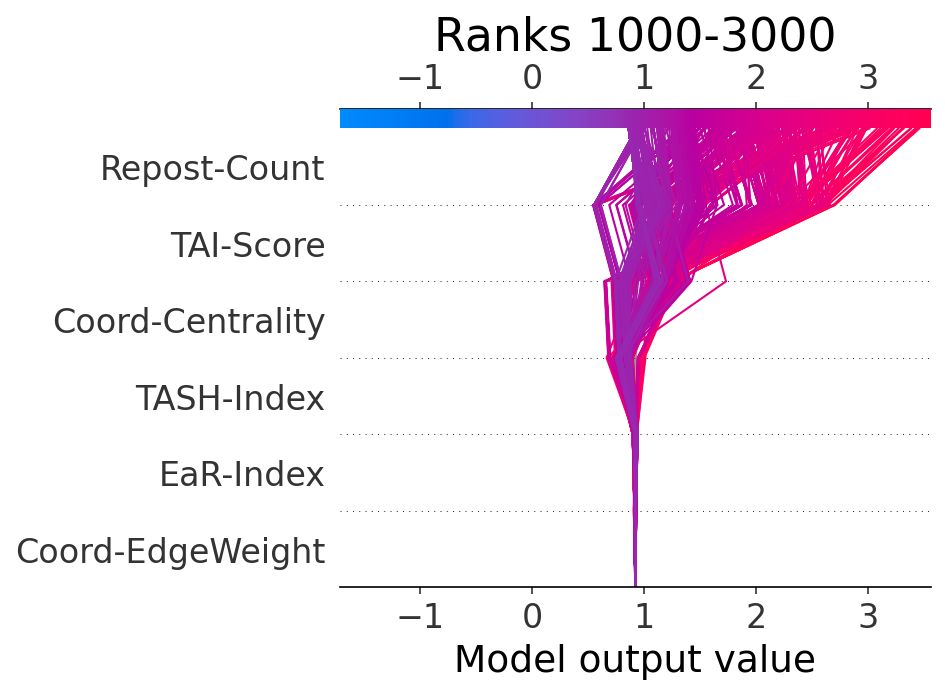}
                \label{fig:sub4}
            \end{subfigure}
        \end{minipage}
    }}
    \caption{\textit{SHAP decision plots} showing how archetypal features contribute to model predictions for users in different ranking intervals. Each line represents a user, starting from the base value and accumulating SHAP values (bottom up) to reach the final prediction on the x-axis. Line color indicates the total SHAP contribution (log-scaled), revealing how the influence of different archetypes varies across the ranking.}
    \label{fig:grid}
\end{figure}

To better understand how the model leverages different features across the ranking spectrum, we analyze SHAP contributions separately for four contiguous intervals in the rank (1–10, 10–100, 100–1000, 1000–3000). This stratified view reveals how the decision process shifts from dominant archetypal profiles to hybrid archetypal profiles as we move down the ranking.

In the top ranks (1–10), the model predominantly relies on the \textit{TASH-Index} and \textit{TAI-Score}, which are the key archetypal features linked to \emph{super-spreader} behavior. This indicates that the highest-ranked users are not merely active in the network, but instead exhibit behaviors that align with the dynamics of influential super-spreaders. 

Between positions 10 and 100, these archetypal features remain important, but features such as \textit{Repost Count} and \textit{Coordination-Centrality} begin to play a stronger role. This suggests that the model is integrating both coordinated and amplifier behaviors, possibly capturing hybrid user types.

In the lower ranks (100--1000), the \textit{Repost Count} emerges as the most influential feature. This reflects the typical behavior of users who primarily engage in amplifying content rather than originating it. While these users may not exhibit high individual impact, their consistent engagement still plays a meaningful role in shaping the overall dynamics of misinformation diffusion—a pattern the model effectively captures.  Notably, the increasing relevance of the \textit{Coordination-Centrality} feature suggests that structured and coordinated activity—rather than isolated individual behavior—becomes a key driver of influence. 
This highlights the importance of coordination signals in identifying actors who contribute collectively to the spread of harmful content, even if their individual metrics are modest. This layered explanation highlights how the model shifts its decision-making across the ranking: features associated with the \emph{super-spreader} archetype --- most notably the TASH-Index --- dominate the top-tier predictions, underscoring the model’s ability to capture key high-risk patterns. As we move down the ranks, signals linked to \emph{amplifiers} (such as Repost Count) and to \emph{coordinated accounts} (Coordination-Centrality) gain relative importance, while other indicators like the EaR-Index (\emph{amplifiers}) and Coord-EdgeWeight (\emph{coordinated accounts}) remain marginal throughout. 

These findings may inform diverse moderation strategies based on distinct behavioral archetypes. Top-ranked users, whose behavior strongly aligns with the super-spreader archetype, might warrant targeted preventive interventions, such as increased moderation scrutiny, fact-checking overlays, or temporary content restrictions. In contrast, mid- and lower-ranked users, whose behavior reflects coordinated or repetitive resharing activity patterns, could be better addressed through broader systemic measures like limiting the amplification of their posts, adjusting recommendation exposure, or applying friction mechanisms (e.g., share limits or warning prompts).

\subsection{Data Scarcity Robustness}\label{subsec:data_scarcity}

To assess the robustness of the most promising methodologies identified in our evaluation (the TASH-Index and the ML approach) in more realistic conditions, we compare their performance in a data-scarce scenario. We aim to answer the following question: \textit{To what extent does delaying the start of data collection impact model performance?}
\begin{figure}[t]
    \centering
    \includegraphics[width=\textwidth]{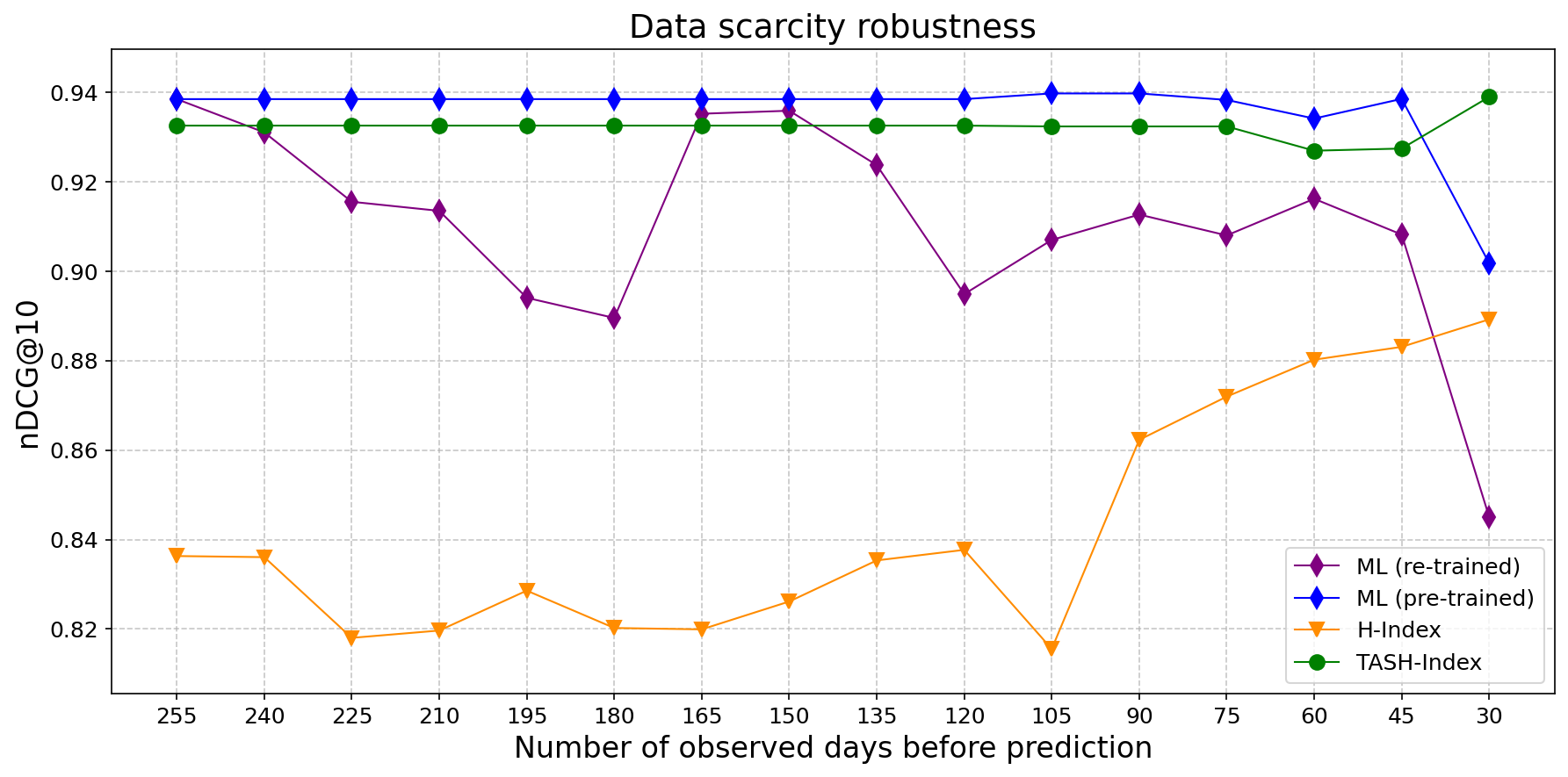}
    \caption{\emph{Performance of the top-performing models under data scarcity conditions.} Each point corresponds to a training of the model with X days of data, obtaining Y as nDCG in evaluation over the same period.}
    \label{fig:data_scarcity_ndcg}
\end{figure}
Data scarcity is a common challenge for many predictive algorithms. The more data a method requires, the longer the waiting time before action can be taken. To address this, we analyze the robustness of the TASH-Index and the machine learning model when faced with limited data.

Figure~\ref{fig:data_scarcity_ndcg} reports the performance of four ranking models under varying data availability conditions. The x-axis shows the number of days of activity used for training, while the y-axis reports the corresponding nDCG@10 score on a fixed test set. Each point represents a complete training and evaluation cycle on a distinct subset of the training data.

The H-Index curve confirms previous results. In this scenario, its performance slightly improves as the observation window shortens, likely because shorter windows capture only the most recent (and thus most predictive) user activity.
The TASH-Index, on the other hand, demonstrates remarkable stability and consistently outperforms the H-Index, indicating its robustness to data scarcity.
The machine learning model was evaluated in two conditions. The \textit{re-trained} ML model, i.e., trained from scratch on each data subset, exhibits instability---its performance fluctuates and degrades as less data becomes available. Conversely, the \textit{pre-trained} ML model, which does not require any re-training, maintains stable and high performance across most of the data reduction range.
Notably, the pre-trained model's performance only starts degrading when trained on the smallest window (30 days). This drop likely occurs because the underlying user archetypal feature vectors become too sparse, impairing the model's ability to make reliable predictions.

\section{Conclusion}\label{sec:conclusion}

\subsection{Discussion}

This study presents a formal taxonomy of three behavioral archetypes of misinformation actors---\emph{super-spreaders}, \emph{amplifiers}, and \emph{coordinated accounts}---and develops corresponding ranking models to assess their influence in the dissemination process. By focusing on observable behavioral patterns rather than user intent, our approach aligns with a core principle of governance: interventions should address systemic harm, regardless of individual motivation. Still, intent-aware approaches could complement our models. For example, integrating influence-based methods such as the one proposed by Zhou et al.~\cite{zhou_this_is_fake_assessing_intent} may enhance ground truth annotations and provide a richer basis for evaluation. Such hybrid models could support more nuanced, tiered moderation strategies that respond differently based on user intent.

A central contribution of this work lies in demonstrating that combining distinct archetypal features within an interpretable model can yield both competitive predictive performance and actionable insights into misinformation dynamics. 

Methodologically, our framework targets behavioral patterns rather than account identity, treating human and automated actors alike. This reflects the increasing overlap between human and bot-like behaviors driven by generative AI, which challenges traditional classification schemes. We argue that behavioral impact offers a reliable signal for moderation, though future work should assess how archetypal behaviors may change with the increasing use of generative AI tools as large language and vision models.

Another core contribution of our work is the TASH-Index, a novel adaptation of the H-Index designed to capture both temporal and structural aspects of misinformation diffusion in a computationally efficient and interpretable manner. Our experiments show that the TASH-Index performs competitively even in low-data settings and enhances the performance of machine learning models when used as a feature. This highlights the value of simple, behavior-based metrics that remain robust and explainable.

From a deployment perspective, our approach is best suited for integration within human-in-the-loop moderation systems, where model-generated rankings act as decision-support tools rather than automated enforcement mechanisms. This allows for scalable interventions, such as prioritizing content for review or limiting visibility, while preserving oversight and accountability. To ensure transparency and reduce potential for misuse, such systems should include safeguards such as auditability and mechanisms for appeal.

\subsection{Limitations}

We acknowledge several limitations of our study. First, the reshare network analyzed captures only direct interactions between original content creators and resharers, without accounting for potential indirect exposure pathways. This limitation stems from platform-level constraints, as most social media APIs do not provide access to complete reshare cascades~\cite{deverna_information_2024}. Second, our labeling of low-credibility content relies on simple yet established techniques based on domain-level assessments. While effective for our purposes, this approach could be extended to other modalities, such as images and videos, contingent on the availability of appropriate credibility annotations.
Third, our ranking models are limited to users observed during training, which restricts their applicability to newly emerging accounts. While the TASH-Index can be updated incrementally with each reshare, enabling real-time applicability, developing analogous, incrementally computable metrics for other archetypal behaviors remains an open research challenge. Finally, our evaluation framework relies on static metrics, which do not account for the dynamic, system-level effects of interventions. Real-world actions, such as content takedowns or account suspensions, can induce behavioral adaptations that reshape the network structure and diffusion processes. Although recent work has begun to explore these dynamics~\cite{Butler2023TheG, truong2024quantifyingvulnerabilitiesonlinepublic, sittar2025agentbasedsimulationsonlinepolitical}, there is still a lack of scalable simulation environments for robust evaluation of intervention strategies. 

\subsection{Future Work}

Several promising research directions arise from this work. First, we aim to further refine the TASH-Index and extend its integration into expressive learning architectures such as graph neural networks, which offer the potential to capture more nuanced temporal and relational structures underlying misinformation diffusion.
Second, we plan to enhance the parameter tuning process of the TASH-Index. Future work will explore adaptive optimization strategies, including gradient-based tuning, Bayesian optimization, or evolutionary algorithms, to enhance generalizability across platforms.
Third, access to richer, multi-platform datasets would enable more comprehensive modeling of user behavior and facilitate the generalization of archetype-based approaches. Such datasets would also support the development of inductive models capable of classifying newly emerging users—an essential step toward real-time moderation pipelines.
Finally, we underscore the importance of simulation-based evaluation frameworks that can model the reactive dynamics of online platforms. These tools are critical for assessing the long-term impact, fairness, and effectiveness of interventions, and for informing the responsible design and deployment of predictive moderation systems.

\section*{Acknowledgements}

This work was supported by the Swiss National Science Foundation through the CARISMA project (Sinergia grant CRSII5 209250). The authors wish to thank Gianluca Nogara for  providing data and supporting the related analysis,
Matt DeVerna for his insightful suggestions, Fil Menczer for offering invaluable insights to this research and all the CARISMA's participants for meaningful discussions.

\appendix

\section{Rankings Methods}\label{appendix_rankings}

This appendix provides additional details on the ranking methods evaluated in our study. We begin with a subsection on the optimization of time-aware approaches, describing how their parameters were tuned. Subsequent subsections present detail ranking methods per each archetype. For each method, we include a visualization of its dismantling track performance, highlighting its effectiveness in reducing misinformation spread over time.

\subsection{Time-Aware Methods Optimization}\label{appendix:optimization}
\begin{figure}[t]
    \centering
    \includegraphics[width=\textwidth]{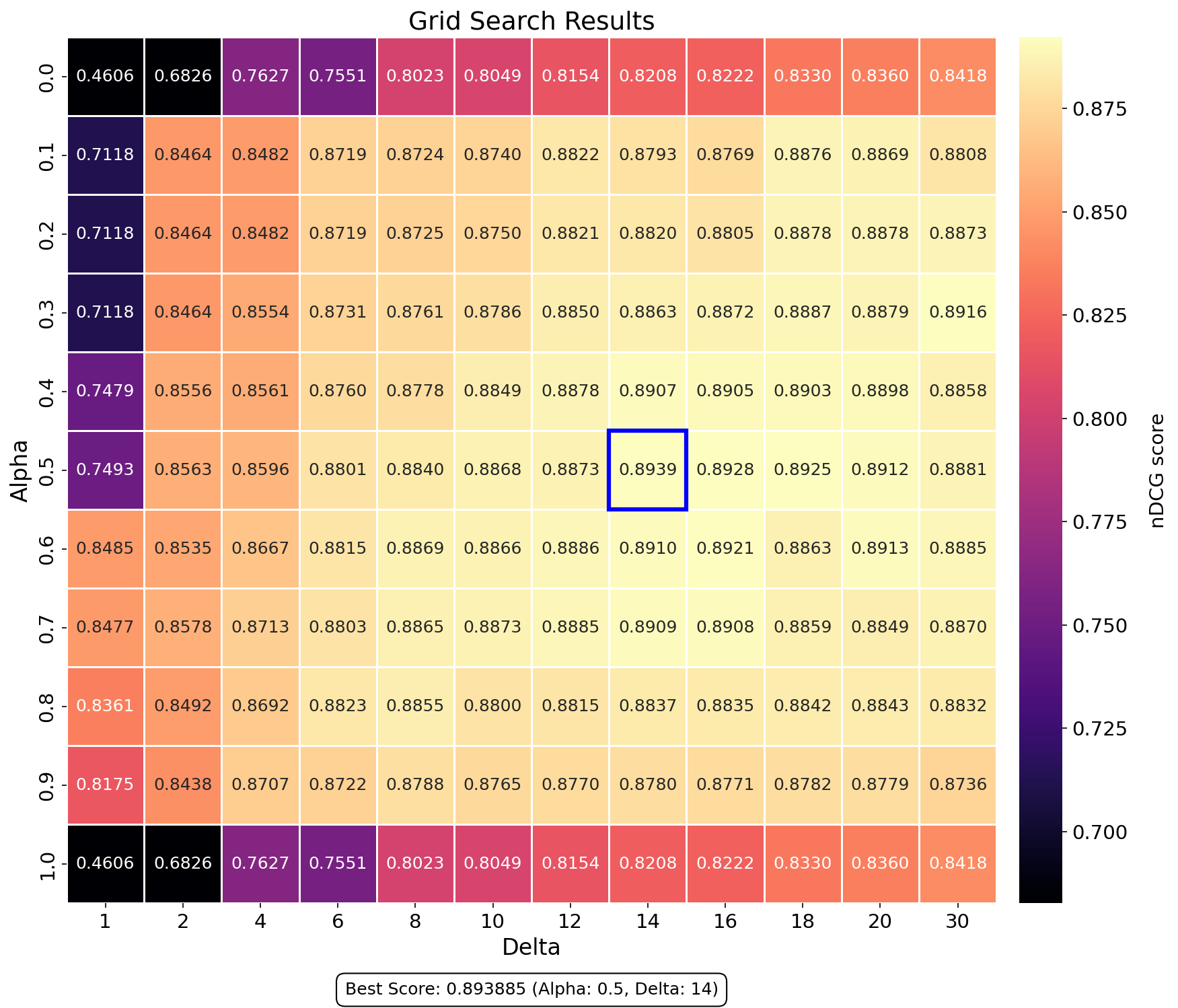}
    \caption{Optimization grid of $\alpha$ and $\delta$ for the TASH-Index using nDCG as the evaluation metric.}
    \label{fig:optimal_search}
\end{figure}

To determine the optimal values of $\alpha$ and $\delta$, we performed a grid search over a predefined range. This resulted in $\alpha=0.5$ and $\delta=14$ days for the TASH-Index model, and $\alpha=0.6$ and $\delta=18$ days for the TAI-index model. A visualization of the search space for TASH-Index is represented in Figure~\ref{fig:optimal_search}.

\subsection{Ranking Amplifiers}\label{appendix:ranking_amplifiers}

\begin{figure}[t]
    \centering
    \includegraphics[width=\textwidth]{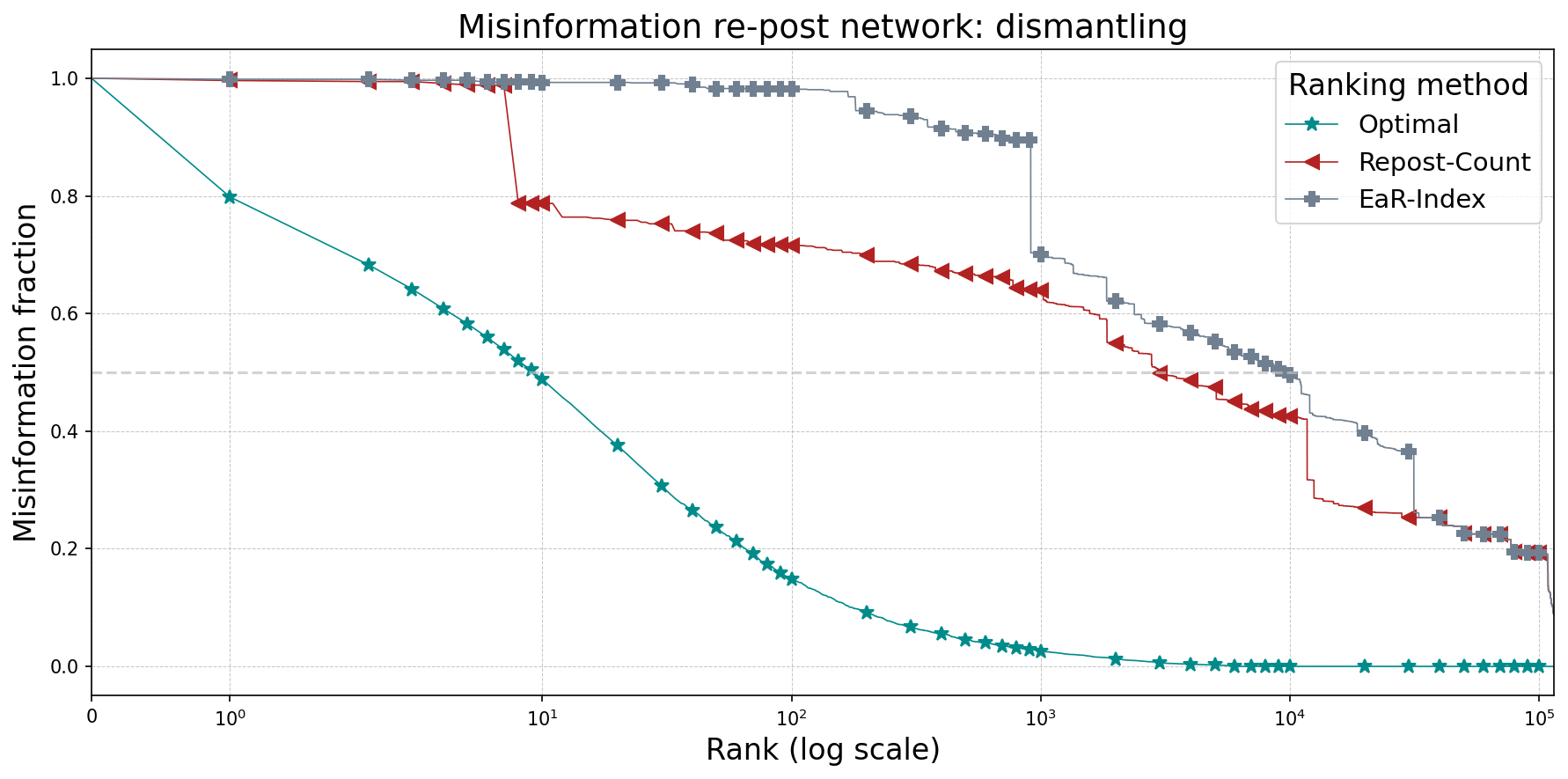}
    \caption{\emph{Performance comparison of \emph{amplifiers}-ranking strategies.}}
    \label{fig:amplifiers_dismantle}
\end{figure}

Figure~\ref{fig:amplifiers_dismantle} evaluates two statistical methods designed to detect \emph{amplifiers}, as introduced in Section~\ref{subsec:amplifier_accounts_rankings}. 

Surprisingly, the \emph{Early Reposter} method—based on the idea that early sharers are influential—performs worse than the simpler \emph{Repost Count} method. This challenges the initial hypothesis that early resharing is a strong proxy for influence in misinformation diffusion.

Figure~\ref{fig:amplifiers_dismantle} shows that reshare frequency is a more effective ranking strategy for identifying \emph{amplifiers} than early reposting. 

\subsection{Ranking Coordinated Accounts}\label{sec:ranking_coordinated}

This subsection evaluates coordinated users using two methods based on the co-reshare similarity network (Section~\ref{subsec:CoordinatedAccounts}). The network is undirected, where edge weights represent cosine similarity between users’ reshare patterns.

\begin{figure}[t]
    \centering
    \includegraphics[width=\textwidth]{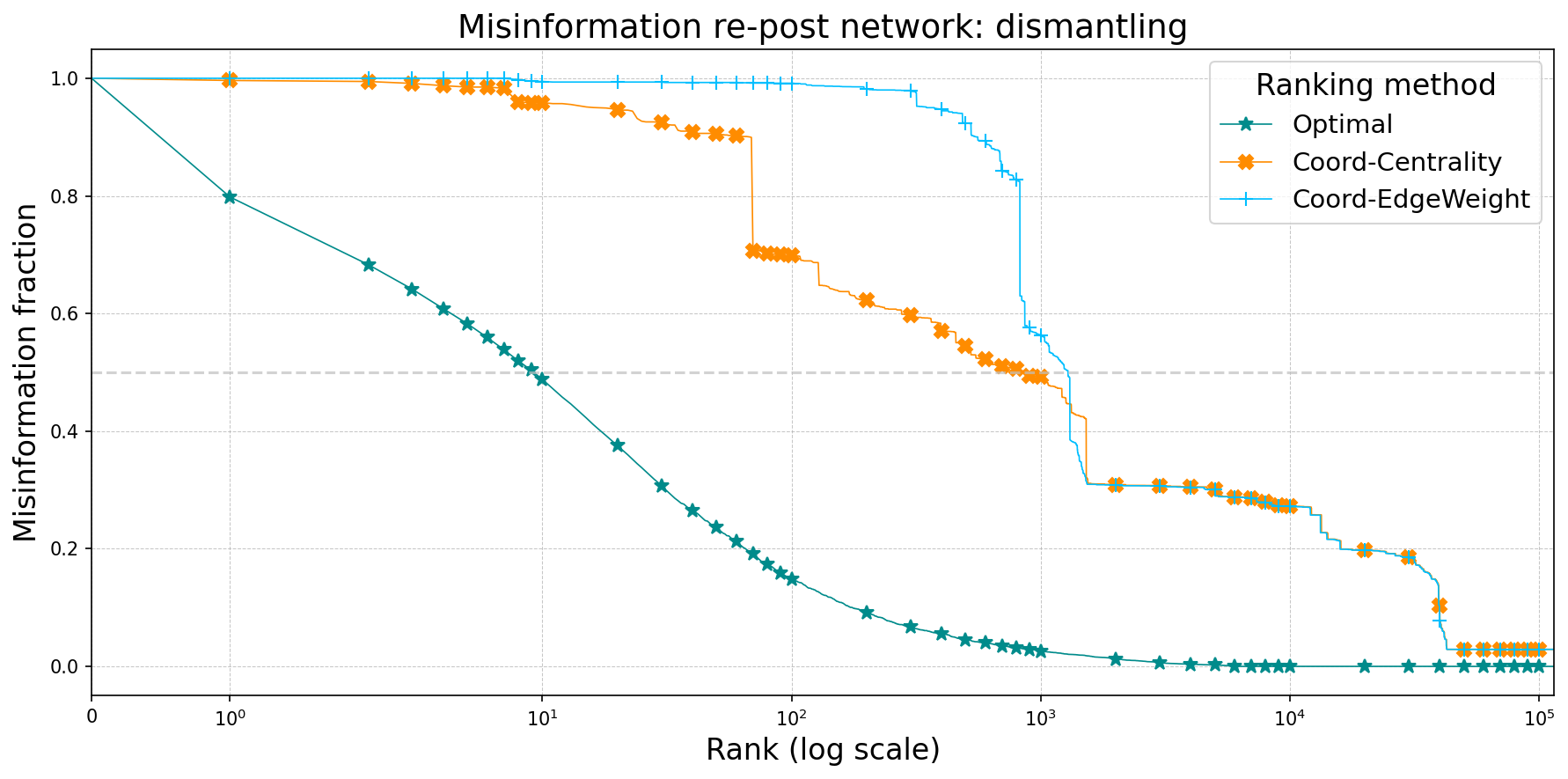}
    \caption{\emph{Performance of coordinated user detection strategies based on similarity networks.}}
    \label{fig:coordinated_misinfo}
\end{figure}

Figure~\ref{fig:coordinated_misinfo} shows the results for:

\begin{itemize}
    \item \emph{Coordination Centrality:} Ranks users by their centrality in the co-reshare similarity network.
    \item \emph{Coordination Max Weight:} Ranks each user by their highest cosine similarity edge.
\end{itemize}

The former method outperforms the latter, suggesting that being embedded in a densely connected coordination cluster is more indicative of influence than having a single strong link. Centrality-based rankings better capture structural importance within coordination networks.

\subsection{Ranking Super-Spreaders}\label{apx:ranking_ss}

\begin{figure}[t]
    \centering
    \includegraphics[width=\textwidth]{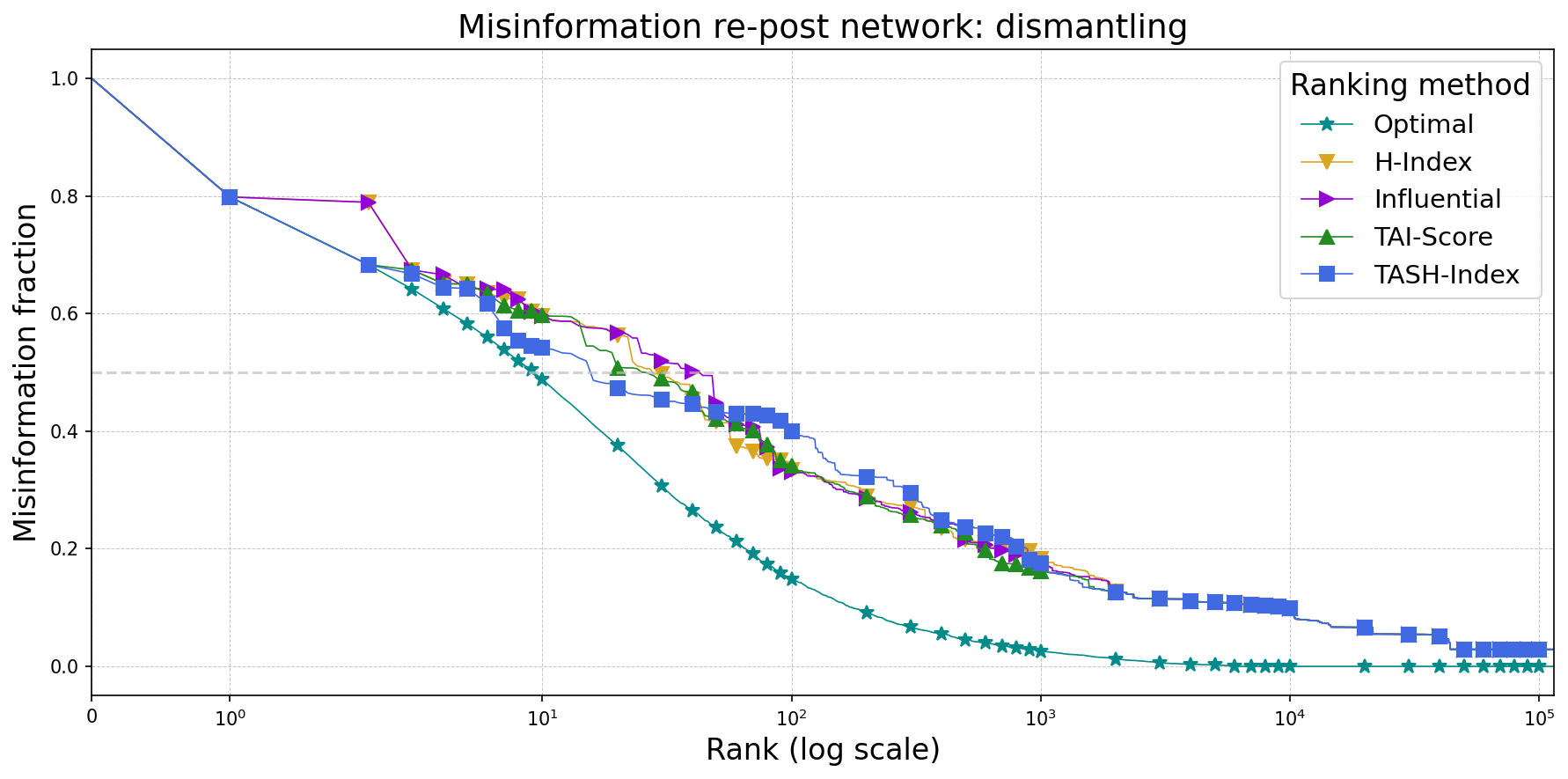}
    \caption{\emph{Comparison of misinformation removal effectiveness across \emph{super-spreader} identification methods.}}
    \label{fig:super_spreaders_misinfo}
\end{figure}

The Figure~\ref{fig:super_spreaders_misinfo} shows the dismantling trajectories of time-aware and time-agnostic methods within the super-spreaders archetype presented in ~\ref{subsec:super_spreaders_ranking}. Both the TASH-Index and TAI-Score show clear improvements when temporal dynamics are included.

\subsection{Ranking Performance in the Multilingual COVID-19 Dataset}\label{ranking_covid_19}

\begin{table}[t]
\centering
\resizebox{\textwidth}{!}{%
\begin{tabular}{lcc|cc|cc|cc|cc}
\toprule
 & \multicolumn{2}{c|}{\textbf{ML}} & \multicolumn{2}{c|}{\textbf{TASH-Index}} & \multicolumn{2}{c|}{\textbf{H-Index}} & \multicolumn{2}{c|}{\textbf{TAI-Score}} & \multicolumn{2}{c}{\textbf{Influence Score}} \\
\cmidrule(lr){2-3} \cmidrule(lr){4-5} \cmidrule(lr){6-7} \cmidrule(lr){8-9} \cmidrule(lr){10-11}
@k  & nDCG  & Quality  & nDCG  & Quality  & nDCG  & Quality  & nDCG  & Quality  & nDCG  & Quality  \\
\midrule
3    & \textbf{0.986}  & \textbf{41.7\%}  & \textbf{0.986}  & \textbf{41.7\%}  & 0.880  & 34.1\%  & 0.914  & 36.2\%  & 0.932  & 37.6\%  \\
5    & \textbf{0.939}  & 42.7\%  & \underline{0.932}  & 42.0\%  & 0.905  & 41.5\%  & \underline{0.932}  & \textbf{43.0\%}  & 0.886  & 38.4\%  \\
10   & 0.918  & 46.8\%  & 0.904  & 45.2\%  & \underline{0.919}  & \textbf{51.3\%}  & \textbf{0.931}  & \underline{49.5\%}  & 0.918  & 49.0\%  \\
20   & \textbf{0.932}  & \textbf{56.9\%}  & \underline{0.926}  & \underline{56.8\%}  & 0.899  & 54.9\%  & 0.925  & 56.7\%  & 0.904  & 54.0\%  \\
100  & \textbf{0.896}  & 68.6\%  & 0.880  & 66.2\%  & 0.872  & 68.4\%  & \underline{0.890}  & 69.5\%  & 0.884  & \textbf{70.7\%}  \\
1000 & \textbf{0.884}  & 81.5\%  & 0.829  & 82.0\%  & 0.823  & 83.7\%  & \underline{0.840}  & 84.8\%  & 0.834  & \textbf{85.1\%}  \\
\midrule
\textbf{Overall} & \textbf{0.926}  & 95.2\%  & \underline{0.886}  & 95.2\%  & 0.876  & 95.2\%  & 0.890  & 95.2\%  & 0.884  & 95.2\%  \\
\bottomrule
\end{tabular}
}
\caption{\emph{Comparison of methods using nDCG and Quality scores at different @k levels on COVID-19 Multilanguage dataset.} The maximum Quality score remains identical across methods because all the nodes are removed from the misinformation reshare network except those that appear after the training phase; consequently, 4.8\% of misinformation cannot be mitigated due to the emergence of new users in the test phase. \textbf{Bold} values denote the best results, \underline{underlined} values indicate the second best.}
\label{tab:ml_ss_multilanguage}
\end{table}

The results on the COVID-19 Multilanguage dataset in
Table \ref{tab:ml_ss_multilanguage} closely mirrors those observed in the VaccinItaly dataset (Table \ref{tab:ml_vs_ss_vaccinitialy}, ensuring the robustness of the findings across different contexts. 
Consistently, the ML model achieves the highest nDCG scores at the top of the ranking, indicating its ability to identify highly influential users. In particular, ML and TASH-index yield the highest nDCG scores, outperforming all other methods, confirming that both models capture strong predictive signals.

\section{Sensitivity Analyses}
This appendix investigates two complementary forms of model sensitivity. The first examines how varying the credibility threshold used to select training data affects model performance. The second explores whether the models tend to prioritize users who consistently share misinformation over those with a mixed content history, offering insight into post-training model behavior.

\subsection{Sensitivity to Credibility Threshold}
We assess how model performance changes when the credibility threshold applied during training is varied. This threshold determines which reshare actions are included in the training data: only actions involving content with credibility scores less than or equal to the threshold are used. For example, a threshold of 20 means that only reshares of content with credibility scores $\leq 20$ are included during training.

Crucially, the test set remains the same and always includes actions labeled using a credibility threshold of 39.0 — which corresponds to content considered \emph{“unreliable because it severely violates basic journalistic standards”}, according to NewsGuard. This fixed reference allows us to isolate the effect of varying the threshold during training.

\begin{figure}[t]
    \centering
    \includegraphics[width=\textwidth]{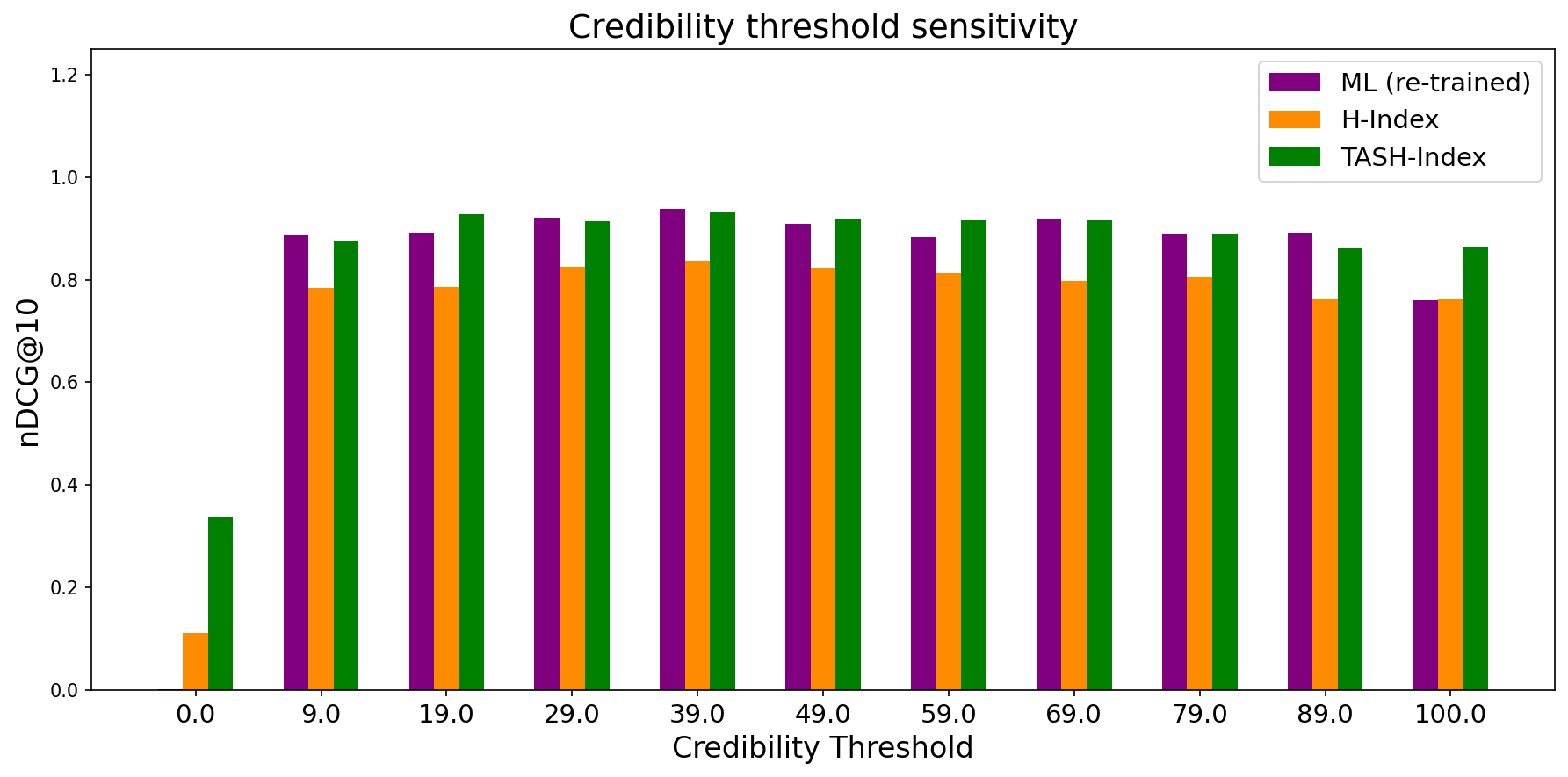}
    \caption{\emph{Impact of varying the credibility threshold during training.} The x-axis indicates the threshold applied, and the y-axis shows model performance (nDCG). For each threshold $x$, only training data involving content with credibility scores $\leq x$ was used.}
    \label{fig:cred_threshold_study}
\end{figure}

As shown in Figure~\ref{fig:cred_threshold_study}, we evaluate three representative models: the H-index approach, the best-performing archetype-based method (TASH-index), and the machine learning model. Overall, results indicate that varying the threshold has limited effects on models performance, with only small fluctuations observed across thresholds.

An exception occurs at the extreme case of a threshold of 0.0. In this setting, only reshare actions involving content with a credibility score of exactly 0 are retained for training. This aggressive filtering results in significantly worse performance, likely due to the sharp reduction in the training data size. 

Interestingly, when the credibility threshold is set to its maximum value (i.e., no filtering is applied and all available data are used for training), the performance of the ML-based model slightly drops, approaching that of the H-index. In contrast, the TASH-index maintains a favorable performance level. 

\subsection{Sensitivity to Credibility in Ranking Composition}\label{appendix_sensitivity}

\begin{figure}[t]
    \centering
    \includegraphics[width=\textwidth]{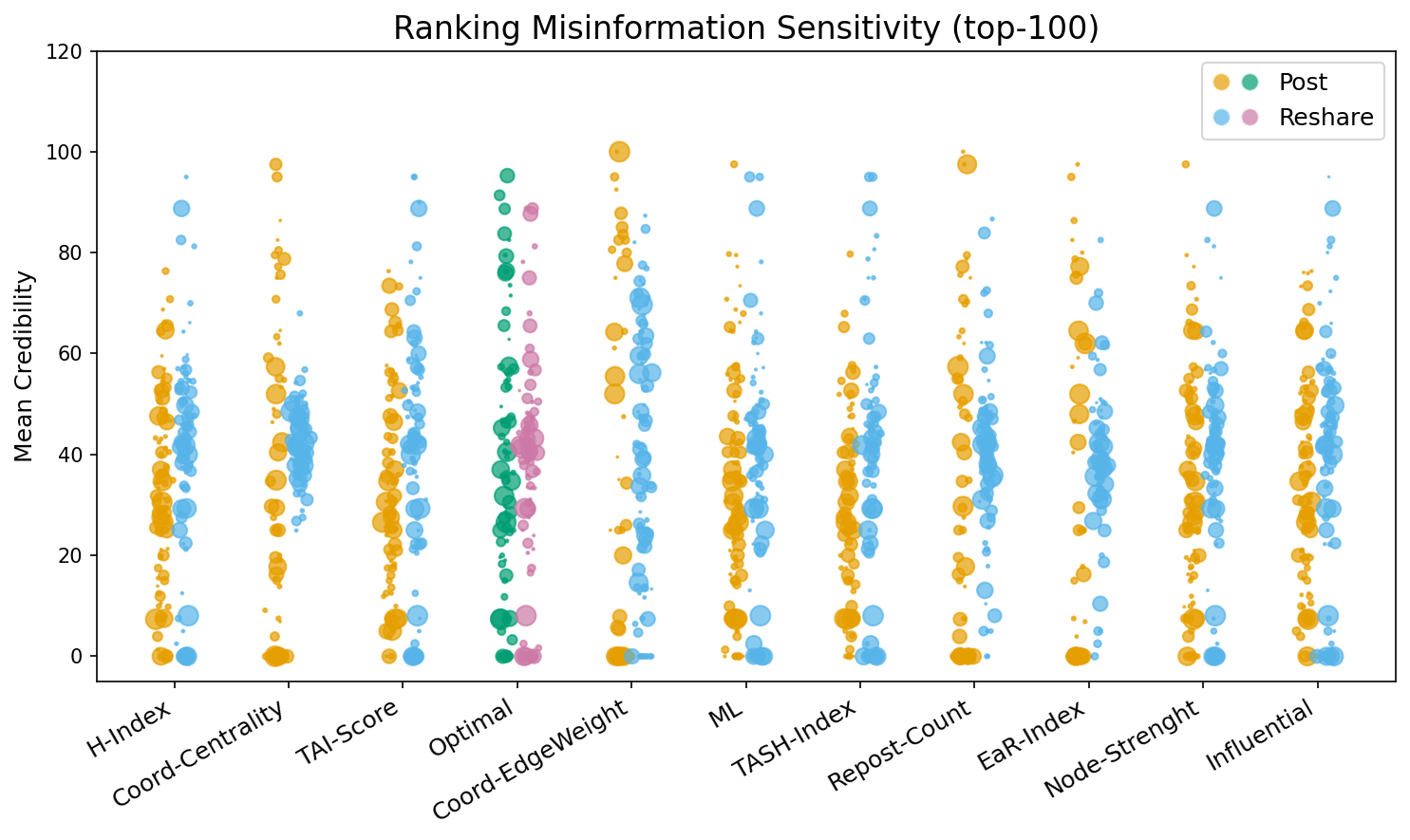}
    \caption{\emph{Credibility profiles of top-100 ranked users across different ranking methods.} Each column represents a ranking method, with users shown as bubbles positioned according to their average credibility scores for original posts (orange) and reshares (blue). For the Optimal ranking, we use distinct colors (green for posts, pink for reshares) to emphasize its role as the ground-truth baseline. Bubble size reflects ranking position, with larger bubbles indicating higher ranks. While all credibility scores are computed from training data, only the Optimal ranking is derived using test-set information based on actual future misinformation spread. This visualization reveals how each method prioritizes users with different credibility profiles, providing insight into the types of actors emphasized by different ranking strategies.}
    \label{fig:misinfo_purity}
\end{figure}

To better understand how different ranking systems prioritize users in terms of the credibility of their content, we perform a sensitivity analysis focused on the top-100 users produced by each method. Figure \ref{fig:misinfo_purity} presents a comparative visualization, where each column corresponds to one ranking method, and each point (bubble) represents a user appearing in the top-100 for that method.

Each user is shown twice: once for their original posts and once for their reshares, distinguished by color: orange for posts and blue for reshares across all rankings, except for the Optimal ranking, which uses green (posts) and pink (reshares) to visually mark its special status.

The vertical axis reports the user’s average credibility score for that specific action type, while the bubble size indicates the user’s position within the ranking (larger bubbles denote higher ranks).

Importantly, only the Optimal ranking (fourth from the left) is computed using future data from the test set: users are ranked based on the actual volume of misinformation they disseminate. This makes it a kind of oracle, offering a behavioral benchmark for comparison. For all methods, including Optimal, credibility scores are always computed from the training data to ensure consistency.

This analysis is not meant to compare performance, but rather to reveal the credibility profiles of the users that each method elevates. Notably, some methods, such as TAI-Score or ML, show wide variability in the credibility of highly ranked users, while others like Coord-EdgeWeight appear to concentrate more around specific ranges. 

\section{Machine Learning}\label{appendix_ml}

This appendix provides additional details on the machine learning methodology used in our study. We describe both the training procedure and the model selection process that guided our final choice. The first subsection covers how the data was split and how model tuning was performed, including the evaluation strategy. The second subsection presents a comparison of different models considered, highlighting their relative performance and justifying the selection of the final model.

\subsection{Model Tuning}\label{appendix_ml_tuning}

All data was chronologically ordered and processed using a sliding window approach, ensuring strict temporal consistency throughout the pipeline (i.e., no leakage of future information into the training data). Importantly, we adopted a \textit{sample-wise} data split, where the split is defined over the number of resharing events (i.e., rows), not over a fixed duration of time (e.g., days or weeks). This choice allows for more granular control over the training set size while preserving the natural temporal order of events.

\begin{figure}[t]
    \centering
    \includegraphics[width=\textwidth]{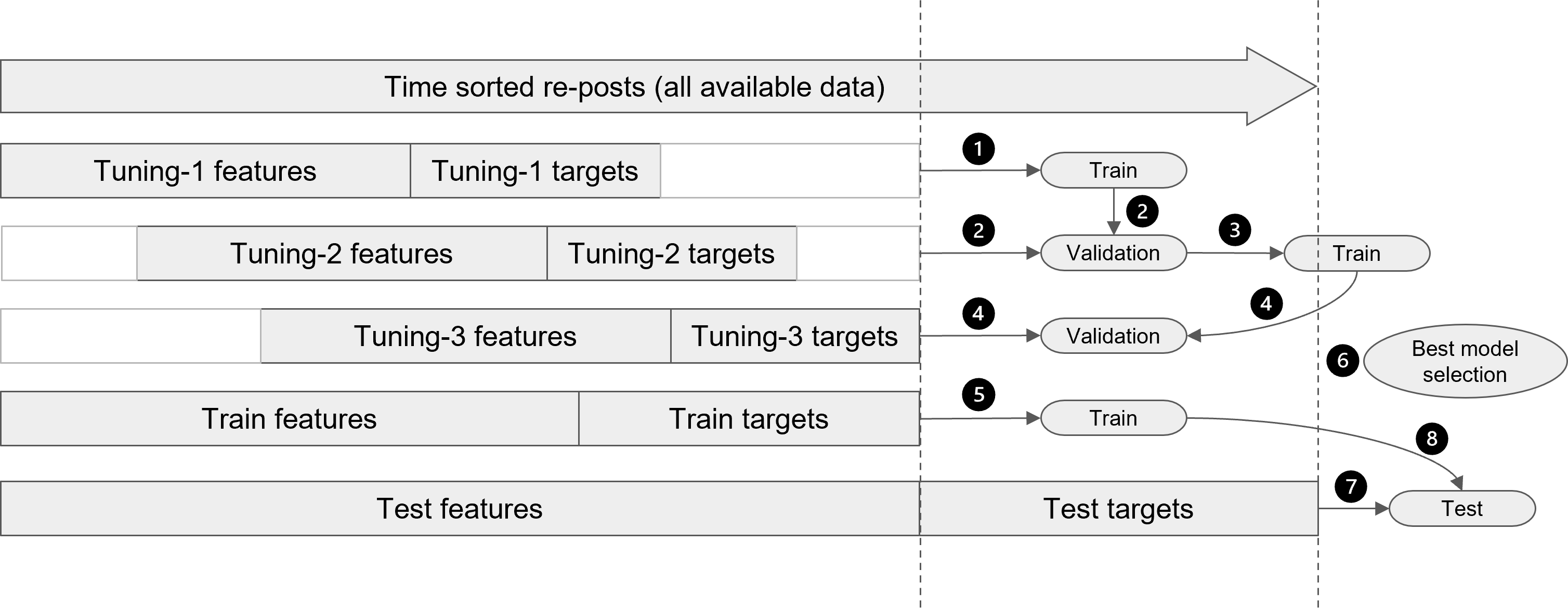}
    \caption{Sliding window procedure for model tuning and evaluation.}
    \label{fig:model_selection_procedure}
\end{figure}

Each sliding window was defined over a contiguous, time-ordered sequence of resharing events. Within each window, models were trained on a prefix of the data and evaluated on the next 20\% of events, simulating forward prediction in real-world deployment settings. Figure~\ref{fig:model_selection_procedure} illustrates this procedure.

This process was repeated across all windows and for every hyperparameter configuration. We selected the configuration that achieved the best average performance across these temporally consistent folds. The procedure ensures robustness to different data regions and guards against overfitting on early or late time periods.

\subsection{Model Selection}\label{appendix_ml_selection}

While user behavior often overlaps across archetypes, machine learning can help disentangle this complexity. We tested multiple models using a hyperparameter grid search, with performance on the test set reported in Table~\ref{tab:model_selection_results}. Given these results, we adopt the Random Forest model as our ML baseline in Section~\ref{subsec:ml_ranking}, particularly to assess how TASH-Index contributes when used alongside other user features.

\begin{table}[H]
    \centering
    \renewcommand{\arraystretch}{1.2}
    \resizebox{\textwidth}{!}{
    \begin{tabular}{lccccccc}
        \toprule
        \textbf{Model} & \textbf{MAE} $\downarrow$ & \textbf{MSE} $\downarrow$ & \textbf{nDCG@3} $\uparrow$ & \textbf{nDCG@5} $\uparrow$ & \textbf{nDCG@10} $\uparrow$ & \textbf{nDCG} $\uparrow$ \\
        \midrule
        Linear Model   & \textbf{1.073}  & 343.34  & 0.955  & 0.920  & 0.851  & 0.890  \\
        Random Forest  & 1.108  & 397.55  & \textbf{0.970}  & \textbf{0.968}  & \textbf{0.939}  & \textbf{0.910} \\
        XGBoost       & 1.079  & 641.62  & 0.750  & 0.860  & 0.850  & 0.864 \\
        Neural Network & 1.287  & \textbf{361.50}  & \textbf{0.970}  & 0.928  & 0.910  & 0.900 \\
        \bottomrule
    \end{tabular}
    }
    \caption{\emph{Performance comparison of tested models on the test set.} Random Forest emerges as the most stable and reliable model overall, especially in managing high target skewness and inter-feature correlation.}
    \label{tab:model_selection_results}
\end{table}

\section*{Declaration of generative AI and AI-assisted technologies in the writing process}
During the preparation of this work the author(s) used DeepL (DeepL SE), ChatGPT (OpenAI) and Claude (Anthropic) in order to enhance the language and readability. After using this tool/service, the author(s) reviewed and edited the content as needed and take(s) full responsibility for the content of the publication.

\bibliographystyle{elsarticle-num} 
\bibliography{misinfo_bibliography}

\end{document}